%% file: main.tex
\documentclass[lettersize,journal]{IEEEtran}
\usepackage{amsmath,amsfonts,amssymb}
\usepackage{tikz}
\usetikzlibrary{shapes.geometric, arrows.meta, positioning}
\usepackage{algorithm}
\usepackage{algpseudocodex}
\usepackage{array}
\usepackage{amsthm}
\usepackage{subcaption}
\usepackage{textcomp}
\usepackage{stfloats}
\usepackage{url}
\usepackage{xcolor}
\usepackage{hyperref}
\usepackage{multirow}
\usepackage{verbatim}
\usepackage{graphicx}
\usepackage{cite}
\usepackage{tikz}
\usepackage{booktabs,threeparttable,pifont}
\usepackage{xcolor} 
\newcommand{\TO}{\textcolor{red!75!black}{TO}}
\newcommand{\ERR}{\textcolor{red!75!black}{ERR}}
\usetikzlibrary{arrows.meta,shapes.geometric,intersections}
     
\usepackage{graphicx}   
\usepackage[utf8]{inputenc}
\usepackage{forest}
\usepackage{enumitem}
\usepackage[nameinlink,capitalize]{cleveref}
\usepackage{pgfplots}
\usepgfplotslibrary{groupplots}
\pgfplotsset{compat=1.18}

\algnewcommand\algorithmicyield{\textbf{yield}}
\algnewcommand\Yield{\algorithmicyield\ }

\algrenewcommand\algorithmicforall{\textbf{parallel for}}
\algrenewcommand\algorithmicfunction{\textbf{worker function}}

\newtheorem{definition}{Definition}
\newtheorem{assumption}{Assumption}

\newtheorem{problem}{Problem}
\newtheorem{proposition}{Proposition}
\newtheorem{lemma}{Lemma}
\newtheorem{corollary}{Corollary}
\newtheorem{theorem}{Theorem}
\newtheorem{remark}{Remark}
\newtheorem{test}{Test}
\theoremstyle{definition} 

\begin{document}

\title{Hyperlyve: Parallelized Enumeration and Exact Lyapunov Stability Certification for Deep ReLU Networks}

\input{heaader}



\maketitle

\begin{abstract}
We introduce Hyperlyve (Hyperplane Partitioned Lyapunov Verification and Enumeration), a framework for proving the asymptotic stability of nonlinear dynamical systems, including those with neural network controllers. Hyperlyve learns and formally certifies continuous piecewise-affine (CPA) neural Lyapunov functions within a counterexample-guided inductive synthesis (CEGIS) loop. Hyperlyve’s key innovation is a parallel algorithm that enumerates the hyperplane arrangement formed by the ReLU activations of deep networks and certifies that Lyapunov conditions hold within each polyhedral region. We derive sound and complete regional tests to verify the asymptotic stability of the exact smooth dynamics (rather than a piecewise-linear approximation) within a compact domain of interest. Furthermore, we formally characterize the computational complexity of the CPA neural Lyapunov verification problem as co-NP-complete, filling a key gap in the literature. Empirical results demonstrate that Hyperlyve enumerates ReLU networks over 150-times faster than competing tools, which contributes to Hyperlyve proving the asymptotic stability of systems that are currently intractable for state-of-the-art verifiers.
\end{abstract}


\section{Introduction}
\IEEEPARstart{C}{yber-physical} systems controlled by a neural network can have excellent performance, yet may exhibit unexpected and sometimes catastrophic behavior.
For such autonomous systems, including vision and foundation models that are used in the control loop, classical stability guarantees remain an essential component of a tiered approach to safety~\cite{robey2026beyond}.
Proving the stability of a closed-loop system with a neural network controller is a well-studied problem, particularly through the lens of reachability analysis~\cite{TNNLSXiangReachbility2, TNNLSXiangReachability1, vincent2025tnnls}.
However, deciding the asymptotic stability of even a simple homogeneous nonlinear system has been shown to be strongly NP-hard~\cite{ahmadi2012difficulty}.

Lyapunov-based asymptotic stability analysis is a cornerstone of control theory. The existence of a Lyapunov function is sufficient to prove the asymptotic stability of a dynamical system. Unfortunately, finding a Lyapunov function is a nontrivial task, as there is no universal method applicable to general nonlinear systems~\cite{khalil2002nonlinear}.
Given the difficulty of hand-deriving Lyapunov functions, a growing body of work learns Lyapunov function as a neural network~\cite{chang2019neural, gaby2022Lyapunov_net, yang2024lyapunov, li2025two, wang2021beta, zhou2022neural, richards2018lyapunov, wu2023neural, petridis2006construction, anon2026hyperlyve, liu2024tool}.
A neural Lyapunov function is generally constructed by implementing a loss function that penalizes violations of the three Lyapunov conditions: zero at the equilibrium, positive definiteness, and decrease along all trajectories~\cite{chang2019neural}.


Although a neural network can learn to satisfy the Lyapunov conditions for a set of sampled points within a domain of interest, this is not a guarantee that the network satisfies the Lyapunov conditions \emph{everywhere} inside the continuous state space. Thus, learning a neural Lyapunov function also requires a formal verification procedure to certify that the stability conditions hold over the entire domain of interest.

Prior work on certifying the validity of a neural Lyapunov function includes \emph{SMT}~\cite{liu2024tool, liu_learning_max_lyapunov, huang_MNLF, edwards2024fossil}, \emph{SOS/SDP}~\cite{liu2024compositionally}, and \emph{MIP/MILP} encodings~\cite{chen2021learning,dai2020counter, dai2021lyapunov}.
Recent work uses the $\alpha,\beta$-CROWN verifier~\cite{zhang2018efficient, wang2021beta} to accelerate verification of smooth neural Lyapunov functions through bound propagation and branch-and-bound optimization~\cite{yang2024lyapunov, li2025two}. However, complete branch-and-bound verification often becomes computationally prohibitive for complex, higher-dimensional systems.

Decomposing a network with ReLU activations into a continuous piecewise-affine (CPA) polyhedral complex by enumerating the hyperplane arrangement formed by the ReLU activation patterns improves the tractability of verifying non-smooth neural Lyapunov functions~\cite{yarbrough2026fitlrp}. Reachable Polyhedral Marching (RPM)~\cite{vincent2025tnnls} enumerates and computes control-invariant sets from activation regions; however, their linear-programming approach relies on piecewise-linear approximations of the dynamics. Similarly, Samanipour et al.~\cite{SamanipourTAC, SamanipouraAllerton, samanipouranalysis} train and enumerate single-layer ReLU neural Lyapunov functions, but also restrict their analysis to piecewise-linear dynamic approximations.

Beyond these dynamical limitations, existing ReLU network hyperplane enumeration methods struggle to parallelize, remaining bottlenecked by per-neuron evaluations~\cite{humayun2022splinecam, villani2024pice, masden2025algorithmic, xu2021traversing, gaines2026characterizing} or network flattening~\cite{vincent2025tnnls, gaines2026characterizing}. Conversely, Berzins' GPU-accelerated approach~\cite{berzins2023edge} is highly parallelized but yields vertices rather than the $H$-representations (half-spaces/facets) necessary for our regional stability checks. Furthermore, while most enumerators assume networks are \emph{generic} (at most $p$ hyperplanes intersecting per vertex in $\mathbb{R}^p$)~\cite{masden2025algorithmic}, CPA neural Lyapunov functions violate this assumption because multiple activation hyperplanes naturally intersect at the origin.

In contrast, we prove the asymptotic stability of \emph{exact} nonlinear, continuous dynamics via parallel decomposition and certification of a CPA neural Lyapunov function.
Our key insight is that the hidden layers of a ReLU network form independent subtrees of activation regions, which can be processed in parallel without a global shared queue. We enumerate the activation regions of non-generic networks via recursive facet march and show that this geometric perspective has better theoretical and empirical performance than the common per-neuron approach. As we decompose the network into a polyhedral complex, we verify the Lyapunov conditions within each region. Tightly constraining verification to each affine region yields a more tractable problem than naively attempting to verify the entire domain at once.

\paragraph{Contributions} Hyperlyve (pronounced ``Hyper-live'') was introduced in~\cite{anon2026hyperlyve} for single-hidden-layer networks. That work focused on \emph{verifying} existing CPA neural Lyapunov functions, and only provided asymptotic soundness under sufficiently dense sampling.
In this work, we improve Hyperlyve to handle deep, multi-layer architectures and integrate it into a CEGIS loop for \emph{Lyapunov function synthesis}, and prove soundness and completeness of the verification algorithm.
The primary contributions of this paper are:
\begin{itemize}
    \item \textbf{Parallel Region Enumeration under Degeneracy:} We introduce a recursive, facet-crossing algorithm tailored to deep ReLU networks that constructs exact $H$-representations without assuming general position, enabling embarrassingly parallel regional execution.
    \item \textbf{Exact Regional Stability Certification:} We establish sound and complete regional criteria to certify asymptotic stability of the true nonlinear dynamics, circumventing the conservativeness of piecewise-linear relaxations.
    \item \textbf{Complexity Characterization:} We prove that the CPA neural Lyapunov verification problem is co-NP-complete, resolving an open theoretical question regarding its computational hardness.
    \item \textbf{End-to-End CEGIS Synthesis \& Empirical Scaling:} We integrate Hyperlyve into a counterexample-guided synthesis pipeline, demonstrating order-of-magnitude verification speedups over state-of-the-art branch-and-bound verifiers and robust precision around origin equilibria.
\end{itemize}

The remainder of this paper is organized as follows. Section~\ref{section:prelim} formalizes the problem, introduces polyhedral preliminaries, and outlines the Hyperlyve framework. Sections~\ref{sec:prove_pos_def} and~\ref{section:verifying_lyap} detail our formal tests for positive definiteness and negative derivative conditions, respectively. Section~\ref{section:region_enum} presents our parallel enumeration algorithm and theoretical complexity bounds. Section~\ref{section:training} presents the CEGIS synthesis algorithm. Section~\ref{sec:experiments} reports experimental results, and Section~\ref{section:conclusion} concludes the paper.

\subsection{Notation}
Let $\mathbb{R}^p$ denote Euclidean space and $\mathcal{C}^n$ the class of functions whose $n$-th derivative is continuous. We denote matrix transposes as $A^\top$, the index set $\{1, 2, \dots, n\}$ as $[n]$, and set cardinality as $|\cdot|$. The $L_\infty$-norm is $\|\mathbf{x}\|_\infty = \max_i |x_i|$. The closure of set $I$ is denoted $\overline{I}$, and $\mathrm{vert}(\cdot)$ extracts the extreme points (vertices) of a polytope. The restriction of a value or function mapping $\mathbf{A}$ to a region $\mathcal{R}$ is written $\mathbf{A}|_{\mathcal{R}}$. The ReLU function is defined as $\sigma(x) = \max(0, x)$.

\section{Problem Statement and Solution Overview}\label{section:prelim}
This section first formalizes the main problem addressed by this paper, namely proving the asymptotic stability of a dynamical system. We then give a concise overview of our solution. Lastly, we provide background information and decompose our solution into the several subproblems that we address throughout this manuscript, including enumerating a neural network into a set of affine regions and verifying that Lyapunov conditions hold across the set.

\subsection{Problem Statement}
Consider the nonlinear continuous-time dynamical system
\begin{equation}
        \dot{\mathbf{x}} = f(\mathbf{x})  = \begin{bmatrix}
            f_1(\mathbf{x}) \\
            f_2(\mathbf{x}) \\
            \vdots \\
            f_p(\mathbf{x})
        \end{bmatrix},
        \label{eq:original_system}
\end{equation}
where $f: \mathbb{R}^p \to \mathbb{R}^p$ is the function representing the system dynamics and $p$ is the dimension of the state space. We treat $f$ as a closed-loop system. These dynamics may be an autonomous system or may include a state-based controller.
Traditionally, the controller would be analytic and provably stable.
However, neural networks are increasingly being trained as controllers and included in closed-loop dynamics,
despite the difficulty of certifying their asymptotic stability~\cite{robey2026beyond}.

\begin{assumption}[Domain of Interest]
    \label{assump:roi}
    The domain of interest $\mathcal{B} \subset \mathbb{R}^p$ is a compact
    convex polytope such that $\mathbf{0} \in \mathcal{B}$.
\end{assumption}

\begin{assumption}[$\mathcal{C}^1$ dynamics]
\label{assump:smooth_dynamics}
    The continuous-time dynamical system given in ~\eqref{eq:original_system} is continuously differentiable, $\mathcal{C}^1$, within $\mathcal{B}$ and has a unique equilibrium at the origin.
\end{assumption}

\noindent The primary goal of this paper is the following:
\begin{problem}[Asymptotic Stability Certification]
    Prove that the nonlinear system~\eqref{eq:original_system}, satisfying Assumptions~\ref{assump:roi} and~\ref{assump:smooth_dynamics}, is asymptotically stable with respect to the origin throughout the domain of interest $\mathcal{B}$.
    \label{problem:main_idea}
\end{problem}

\subsection{Preliminaries}
For our purposes, a ReLU neural network $V:\mathbb{R}^p\to\mathbb{R}$ is a fully connected feedforward network with $L$ hidden layers of $N_\ell$ neurons each, indexed by $\ell \in [L]$.
The network $V$ represents a continuous piecewise affine (CPA) function formed by the composition of affine maps induced by the ReLU activations of its hidden neurons. Each neuron is either \emph{inactive} when its pre-activation input is negative or \emph{active} when its pre-activation input is positive.
The switching surface between these two states is a hyperplane along the zero-level set.
These hyperplanes form an arrangement that partitions the input space $\mathbb{R}^p$ into a complex of convex polyhedra~\cite{montufar2014number}, within each of which $V$ restricts to a single affine function.
We exploit this geometric structure in our verification procedure.

\begin{definition}[Arrangement of Hyperplanes]~\cite{stanley2007introduction}
Let $H_n = \{\mathbf{x} \in \mathbb{R}^p \mid \mathbf{w}_n^T \mathbf{x} + b_n = 0\}$ represent a hyperplane in $\mathbb{R}^p$, where $\mathbf{w}_n \in \mathbb{R}^p \setminus \{\mathbf{0}\}$ and $b_n \in \mathbb{R}$ for $n \in [N_\ell]$. Then $\mathcal{A} = \{H_1, H_2, \dots, H_{N_{\ell}}\}$ is an arrangement of $N_{\ell}$ hyperplanes. 
\label{def:arrangement}
\end{definition}
The arrangement of hyperplanes for each layer defines the $H$-representation (intersection of half spaces) of affine regions. For deep ReLU networks, the weights and biases of the $i$-th hidden layer are a distinct hyperplane arrangement, $\mathcal{A}_i$.

\begin{definition}[Open Connected Region]~\cite{stargalla2025computational}
    Given a ReLU network $V$, an \emph{open connected region} of $V$, denoted $\mathcal{U}$, is an open inclusion-wise maximal connected subset of $\mathbb{R}^p$ on which $V(\mathbf{x})$ is affine.
    \label{def:open_region}
\end{definition}

\begin{definition}[Local Activation Region] \label{def:region}
    Let V be a ReLU network with corresponding open connected regions $\{\mathcal{U}_t\}_{t = 1}^{T}$, $T < \infty$. The \emph{local activation regions} are defined as the intersection of each $U_t$ with $\mathcal{B}$, and is denoted $\mathcal{R}_i$, $i \in [M]$.  
\end{definition}

On each region $\mathcal{R}$, the neural network $V$ restricts to a single affine function $\mathbf{x} \mapsto \nabla V^\top|_R\mathbf{x} + b_R$.
We use the terms ``region'' and ``activation region'' interchangeably with local activation region.
Every region is bounded by \emph{facets}, which are key elements for enumerating the regions of $V$.
An \emph{activation region} is so called because crossing a facet of an activation region switches some neuron from active to inactive. 

\begin{definition}[Facet] \label{def:facet}
Let $\mathcal{R}$ be an open region of a hyperplane arrangement $\mathcal{A}$, and let $\overline{\mathcal{R}}$ denote its closure. A facet of $\mathcal{R}$ is a $(p-1)$-dimensional face of $\overline{\mathcal{R}}$, given by $F = H \cap \overline{\mathcal{R}}$ for some $H \in \mathcal{A}$ such that $\dim(H \cap \overline{\mathcal{R}}) = p - 1$. The set of all facets of $\mathcal{R}$ is denoted by $\text{facets}(\mathcal{R})$.
\end{definition}

Geometrically, the facets of $\mathcal{R}$ correspond to the non-redundant linear constraints defining its $H$-representation. We employ facets both in our enumeration of $V$ and our proof of asymptotic stability.

\begin{definition}[Weak Lyapunov function]\label{def:weak_lyap}
    A continuous function $V : \mathcal{B} \to \mathbb{R}$ is a \emph{weak Lyapunov function} over a compact domain $\mathcal{B} \subset \mathbb{R}^p$ with $\mathbf{0} \in \mathcal{B}$ for the system~\eqref{eq:original_system} if it satisfies the conditions:
    \begin{enumerate}[label=\textbf{\arabic*.}, ref=\arabic*]
        \item \textbf{Positive-Definiteness:}\label{cond:pos_def}
        \begin{enumerate}[label=(\alph*), ref=\theenumi(\alph*)]
            \item $V(\mathbf{0}) = 0$\label{cond:zero_origin}
            \item $V(\mathbf{x}) > 0$ for all $\mathbf{x} \in \mathcal{B}\setminus \{\mathbf{0}\}$\label{cond:positivity}
        \end{enumerate}
        
        \item \textbf{Decrease:}\label{cond:dec}
        $D_f^+V(\mathbf{x}) \leq 0$ for all $\mathbf{x} \in \mathcal{B}\setminus \{\mathbf{0}\}$, where the upper Dini derivative is represented by
        \[
            D_f^+V(\mathbf{x}) = \limsup_{\delta\to 0^+}\frac{V(\mathbf{x} + \delta f(\mathbf{x})) - V(\mathbf{x})}{\delta}.
        \]
    \end{enumerate}
\end{definition}

Standard Lyapunov literature often defines Condition~\ref{cond:dec} using the Lie derivative $\nabla V(\mathbf{x}) \cdot f(\mathbf{x})$, which requires continuously differentiable ($C^1$) functions and smooth activations (e.g., $\tanh$). Since ReLU networks are continuous piecewise affine (CPA) rather than smooth, we employ the generalized upper Dini derivative~\cite{moulay2010some}. For continuously differentiable regions, $D_f^+V(\mathbf{x})$ coincides exactly with $\nabla V(\mathbf{x}) \cdot f(\mathbf{x})$.

\subsection{The Hyperlyve Framework}

\input{cegis}

To solve Problem~\ref{problem:main_idea}, we introduce \emph{Hyperlyve} to synthesize and formally verify a CPA neural Lyapunov function $V(\mathbf{x})$. As illustrated in Figure~\ref{fig:cegis_loop}, Hyperlyve operates within a counterexample-guided inductive synthesis (CEGIS) loop composed of two primary modules:
\begin{enumerate}
    \item \textbf{Learner:} Trains a candidate Lyapunov network $V(\mathbf{x})$.
    \item \textbf{Hyperlyve:} Formally verifies $V$ over the entire continuous domain $\mathcal{B}$ through a two-stage parallel pipeline:
    \begin{itemize}
        \item \emph{Region Enumeration (Section~\ref{section:region_enum}):} enumerates the activation regions of $V$ via a hierarchical facet-crossing algorithm, decomposing $\mathcal{B}$ into a polyhedral complex.
        \item \emph{Regional Certification (Sections~\ref{sec:prove_pos_def}--\ref{section:verifying_lyap}):} tests the Lyapunov conditions over each activation region of the complex. Testing vertices is sufficient for positive-definiteness. Since $V$ is affine within $\mathcal{R}$, the decrease condition simplifies to $\nabla V \cdot f(\mathbf{x})$, which is certified on the exact nonlinear dynamics using regional branch-and-bound.
    \end{itemize}
\end{enumerate}

Hyperlyve returns a counterexample for any region that fails a Lyapunov condition, and these counterexamples are used for further training of $V$. If no counterexamples exist, then Hyperlyve has found a Lyapunov function for~\eqref{eq:original_system}, which proves the system is asymptotically stable and answers Problem~\ref{problem:main_idea}.

The polyhedral decomposition provides a decisive computational advantage over holistic verifiers.
Standard complete verifiers (e.g., branch-and-bound in $\alpha,\beta$-CROWN) bound the composition of the Lyapunov network $V$ and nonlinear dynamics $f$ simultaneously. Across the entire domain $\mathcal{B}$, these coupled bounds loosen rapidly, triggering extensive branch splitting.
In contrast, by explicitly reasoning over each affine region $\mathcal{R}$, Hyperlyve eliminates the nonlinearity of $V$ for each verification subproblem. Since the gradient of $V$ is constant within each region, certifying the Dini derivative reduces to proving $\nabla V|_\mathcal{R} \cdot f(\mathbf{x}) < 0$, where the nonlinearity is exclusively due to the closed-loop dynamics.

\begin{figure}[t]
    \centering
    \includegraphics[width=0.9\linewidth]{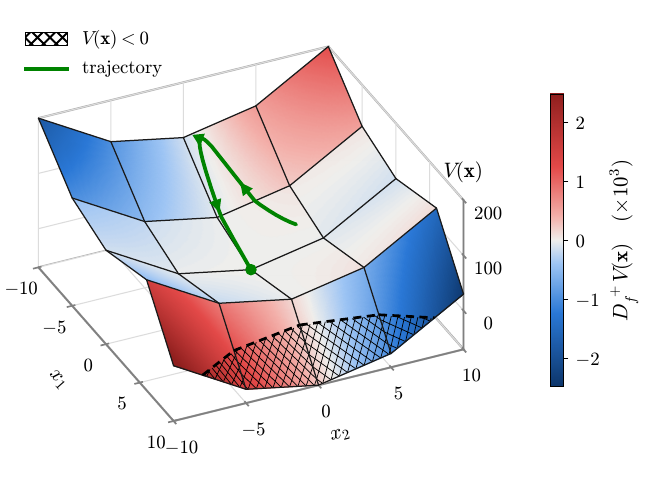}
    \caption{A candidate ReLU Lyapunov function ($8$ neurons in $1$ hidden layer) attempting to certify stability for the Ahmadi-Krstic-Parrilo (AKP) dynamics~\cite{ahmadi2011globally} on $\mathcal{B} = [-10, 10]^2$. Trajectories converge asymptotically to the origin (green path), but the candidate $V$ fails both positive-definiteness (hatched) and Dini-derivative decrease (shaded red). Hyperlyve decomposes the domain into sixteen convex regions and provides counterexamples for the ten regions with violations so that $V$ can continue to be trained.}
    \label{fig:v3d_akp}
\end{figure}

\noindent\textbf{Motivating Example.}
Figure~\ref{fig:v3d_akp} demonstrates regional decomposition and verification on the Ahmadi-Krstic-Parrilo (AKP) polynomial system ($\dot{x}_1 = -x_1 + x_1 x_2$, $\dot{x}_2 = -x_2$)~\cite{ahmadi2011globally}. AKP is known to not have any polynomial Lyapunov function, so we choose to prove stability with a CPA neural Lyapunov function. A candidate ReLU network partitions the domain into 16 convex activation regions. Hyperlyve evaluates each region independently: checking the vertices of each region verifies positive definiteness, while branch-and-bound certifies the Dini derivative on the continuous dynamics. In this example, Hyperlyve identifies violations in ten of the sixteen regions, producing localized counterexamples that guide the CEGIS loop to synthesize a valid network.

\subsection{Subproblems}
The core contribution of Hyperlyve is a sound and complete framework that identifies counterexamples violating Definition~\ref{def:weak_lyap}. These counterexamples drive a CEGIS loop to discover and formally certify CPA neural Lyapunov functions for nonlinear dynamical systems.

\begin{assumption}
    \label{assump:given_v}
    We are given a feedforward ReLU neural network $V: \mathbb{R}^p \to \mathbb{R}$ with $L$ hidden layers of width $N_\ell$, a bounded domain of interest $\mathcal{B} \subset \mathbb{R}^p$ satisfying Assumption~\ref{assump:roi} and continuous dynamics $f$ adhering to Assumption~\ref{assump:smooth_dynamics}. Also, $V$ partitions $\mathcal{B}$ into activation regions, denoted $\mathcal{R}_i$, $i \in [M]$.
\end{assumption}

To formally prove stability (Problem~\ref{problem:main_idea}), Hyperlyve solves the following decision problems.
\begin{problem}\label{prob:pos_def}
Under Assumption~\ref{assump:given_v}, is $V$ positive-definite (Condition~\ref{cond:pos_def}) on $\mathcal{B}$?
\end{problem}

\begin{problem}\label{prob:dec}
Under Assumption~\ref{assump:given_v}, does a positive-definite $V$ satisfy the decrease condition (Condition~\ref{cond:dec}) for the system dynamics $f(\mathbf{x})$ on $\mathcal{B}$?
\end{problem}

Solving Problems~\ref{prob:pos_def} and~\ref{prob:dec} requires conducting verification tests subject to the linear inequality constraints that define each affine region of $V$. These constraints must be discovered, which requires us to enumerate the ReLU activation hyperplanes of $V$, and giving us an additional geometric problem.

\begin{problem} \label{prob:enumerate}
Extract the exact $H$-representation of the polyhedral decomposition induced by $V$ on $\mathcal{B}$.
\end{problem}
Problem~\ref{prob:pos_def} and~\ref{prob:dec} are addressed in Section ~\ref{sec:prove_pos_def} and~\ref{section:verifying_lyap}, respectively. Section~\ref{section:region_enum} solves Problem~\ref{prob:enumerate} and also presents the Hyperlyve verification algorithm. The integration of Hyperlyve into a CEGIS loop is discussed in Section~\ref{section:training}.

\section{Positive-definite Condition}
\label{sec:prove_pos_def}

Recall that Condition~\ref{cond:pos_def} for a weak Lyapunov function is its positive-definiteness. In this section we provide a formal test for both components of this condition. We also show positive-definite determination is co-NP-complete (see Theorem~\ref{thm: pos_np}).

\subsection{Tests for Positive-definiteness}
Note that testing if a ReLU network is strictly positive within $\mathcal{B}$ can be well-accomplished with reachability analysis~\cite{bak2021nfm}; however, we introduce a novel approach because $V$ must simultaneously be exactly zero only at the origin and positive elsewhere within $\mathcal{B}$, which is a different formulation than what reachability analysis currently supports.

Condition~\ref{cond:zero_origin} is a trivial test at the origin: 
\begin{test}[Zero at Origin — Condition~\ref{cond:zero_origin}]
    \label{test:cond:zero_origin}
    Evaluate $V(\mathbf{0})$. If $V(\mathbf{0}) \neq 0$, $\mathbf{x}^*=\mathbf{0}$
    is a counterexample. Otherwise, Condition~\ref{cond:zero_origin} is satisfied.
\end{test}

Hyperlyve checks the strict positivity of $V$ elsewhere within $\mathcal{B}$ by checking the vertices of the polyhedral decomposition.
\begin{proposition}
\label{prop:positive_check}
     Suppose Assumptions~\ref{assump:roi}, ~\ref{assump:smooth_dynamics}, and ~\ref{assump:given_v} hold.  Consider an arbitrary activation region $\mathcal{R}$. If there exists $\mathbf{x}_0 \in \mathcal{R}$ such that $V(\mathbf{x}_0) \leq 0$, then there exists a vertex $\mathbf{a}$ of $\overline{\mathcal{R}}$ such that $V(\mathbf{a}) \leq 0$.
\end{proposition}

\begin{proof}
    We proceed by contradiction. Suppose that there exists some $\mathbf{x}_0 \in \mathcal{R}$ such that $V(\mathbf{x}_0) \leq 0$. Further, suppose that $V$ is positive at all vertices. 

    It is well-known that $V$ is affine over $\mathcal{R}$. By the continuity of $V$, $V$ is affine in $\overline{\mathcal{R}}$. Then, by the Fundamental Theorem of Linear Programming \cite{dantzig2016linear}, it is known that a minimum occurs at some vertex~$\mathbf{a}$. Therefore, $V(\mathbf{x}_0)$ is larger than the minimum, and $V(\mathbf{a}) \leq V(\mathbf{x}_0) \leq 0$. Since we assumed $V(\mathbf{a}) > 0$, we have a contradiction.   
\end{proof}

A consequence of the above proposition is that if there exists a counterexample to the positive definiteness condition in the interior of a linear activation region, then a counterexample must exist at a vertex of the same region. Thus, positive definiteness of the candidate neural network can be determined by evaluating $V$ at each of the vertices, which can be done sequentially in a finite number of steps. 
\begin{test}[Positivity — Condition~\ref{cond:positivity}]
    \label{test:cond:positivity}
    Consider an activation region $\mathcal{R}$. If there exists a vertex $\mathbf{a} \in \bigcup_i \mathrm{vert}(\overline{\mathcal{R}}) \setminus \{\mathbf{0}\}$
    where $V(\mathbf{a}) \leq 0$, $\mathbf{a}$ is a counterexample.
    Otherwise, Condition~\ref{cond:positivity} is satisfied.
\end{test}

\begin{remark}
    Test~\ref{test:cond:positivity} can equivalently be conducted by solving a linear program (LP)~\cite{SamanipourTAC, SamanipouraAllerton}.
    However, as we shall show, our solution to Problem~\ref{prob:dec} incidentally yields the vertices of each region, and a batched forward pass of vertices through $V$ is generally faster than solving a set of LPs.
\end{remark}

\begin{corollary}
\label{corollary:vertex_origin}
        Suppose Assumptions~\ref{assump:roi}, ~\ref{assump:smooth_dynamics}, and ~\ref{assump:given_v} hold. If $V$ satisfies Condition~\ref{cond:pos_def} then the origin lies exactly on a vertex of an activation region. 
\end{corollary}
\begin{proof}
    Suppose that the origin is not a vertex of any activation region. Since the origin is not at a vertex, it cannot be a minimum of $V$. Thus, there exists an $\mathbf{x} \in \mathcal{B}$ such that $V(\mathbf{x}) < V(\mathbf{0}) < 0$, which violates Condition~\ref{cond:positivity}. This completes the proof. 
\end{proof}

The above corollary gives insight into the geometric structure of ReLU networks that satisfy Tests~\ref{test:cond:zero_origin} and~\ref{test:cond:positivity}. Specifically, Corollary~\ref{corollary:vertex_origin} suggests that the corresponding hyperplane arrangements of most CPA Lyapunov functions are \emph{not} in general position (worst-case scenario in terms of maximizing activation regions), since it is expected that many hyperplanes intersect at the origin. This notion will resurface in Section~\ref{section:region_enum}.  
We now formally show that these tests can verify a candidate Lyapunov function, completing our solution to Problem~\ref{prob:pos_def}. 

\begin{lemma}\label{lemma: answer_prob2}
    Suppose Assumptions~\ref{assump:roi}, ~\ref{assump:smooth_dynamics}, and ~\ref{assump:given_v} hold. If every activation region $\mathcal{R}_i \in \mathcal{B}$ successfully passes Tests~\ref{test:cond:zero_origin} and~\ref{test:cond:positivity}, then $V$ is positive-definite.
\end{lemma}
\begin{proof}
    Define the following function with compact support: 
    \begin{equation}
        g_i(\mathbf{x}) = \begin{cases}
    V(\mathbf{x}) & \text{if } \mathbf{x} \in \overline{\mathcal{R}}_i \\
    0   & \text{if } \mathbf{x} \in \mathcal{B} \setminus \overline{\mathcal{R}}_i. \\
\end{cases}
    \end{equation}
    By the success of Test~\ref{test:cond:positivity}, each $g_i$ is positive in the respective $\overline{\mathcal{R}}_i$, except for the functions whose support has a vertex at the origin by Corollary~\ref{corollary:vertex_origin}. Note that $\sum_{i = 1}^M g_i(\mathbf{x}) = V(\mathbf{x})$. By Test~\ref{test:cond:zero_origin}, each $g_i$ is zero at the origin, and thus $V$ is also zero at the origin. Furthermore, each $g_i$ is positive (except for regions that contain the origin) over disjoint activation regions, which implies that $V$ is also positive everywhere excluding the origin. Thus, $V$ is a positive-definite CPA function. 
\end{proof}

\subsection{Complexity of Proving Positive-definiteness}

In general, it has been shown that verifying properties of ReLU networks is NP-complete~\cite{katz2017reluplex, salzer2023reachability}.
However, to the best of our knowledge, no formal proof of the complexity of proving the positive-definite condition of a CPA neural network has been given, so we include it here.

\begin{theorem}\label{thm: pos_np}
    Problem~\ref{prob:pos_def} is co-NP-complete.
\end{theorem}
\begin{proof}
We formally define the decision version of Problem~\ref{prob:pos_def} as the conjunction of two conditions, $T = A \land B$, where $A$ denotes the predicate $V(\mathbf{0}) = 0$ and $B$ denotes the predicate $V(\mathbf{x}) > 0$ for all $\mathbf{x} \in \mathcal{B} \setminus \{0\}$.

The complement, $\neg T$, is given by the disjunction $\neg A \lor \neg B$. Note that $\neg T$ is true if $V(\mathbf{0}) \neq 0$ or if there exists an $\mathbf{x} \in \mathcal{B} \setminus \{0\}$ such that $V(\mathbf{x}) \leq 0$.

Since $A$ can be solved in polynomial time, $\neg A$ is also in P. It has been shown that determining $\neg B$ is NP-complete~\cite{murty1987some}. Since $\neg T$ is the disjunction of a problem in P and an NP-complete problem, $\neg T$ is NP-complete. Consequently, the original decision problem $T$ is co-NP-complete.
\end{proof}

It is interesting to note that if the vertices for each activation region are given as an input parameter, Problem~\ref{prob:pos_def} is in $P$ (see Tests~\ref{test:cond:zero_origin} and~\ref{test:cond:positivity}). However, the previous theorem shows that the problem is co-NP-complete if the vertices are not provided. Thus, the computational complexity of our method is dominated by the task of finding the activation regions and the corresponding vertices, which is discussed in Section~\ref{section:region_enum}.

\section{Decrease Condition}
\label{section:verifying_lyap}

In this section, we present how Hyperlyve proves the asymptotic stability of a given dynamical system~\eqref{eq:original_system} by certifying that a positive-definite network $V$ (see Section~\ref{sec:prove_pos_def}) decreases along all trajectories of the dynamics within $\mathcal{B}$ (Condition~\ref{cond:dec}), so is thus a weak Lyapunov function (Definition~\ref{def:weak_lyap}).
As in the previous section, we first provide a per-region test and then show that the problem is co-NP-complete.

\subsection{Proving Asymptotic Stability}
The purpose of verifying a neural Lyapunov function is to prove asymptotic stability for a nonlinear dynamical system. The next result demonstrates that a positive-definite network can be shown to be asymptotically stable by only verifying that the Lie derivative is strictly negative in the interior of each of the linear activation regions; explicitly checking the boundaries between regions is unnecessary.

\begin{theorem}
\label{thm: interior_ver} Suppose Assumptions~\ref{assump:roi}, ~\ref{assump:smooth_dynamics}, and ~\ref{assump:given_v} hold. 
    Consider the dynamical system in~\eqref{eq:original_system} and a positive-definite network $V: \mathcal{B} \to \mathbb{R}_{\geq 0}$. If $D_f^+V(\mathbf{x}) < 0$ for all $\mathbf{x} \in \mathcal{R}_i$ for each $i$, then $V$ is a weak Lyapunov function.
\end{theorem}

\begin{proof}
    Appealing to Assumption~\ref{assump:roi}, the origin is contained in $\mathcal{B}$. Recall a positive-definite $V$ satisfies Lyapunov Condition~\ref{cond:pos_def}. Thus, $V$ is zero at the origin and is positive everywhere within $\mathcal{B}$ excluding the origin. 

    By Assumption~\ref{assump:given_v}, $\mathcal{B} = \bigcup_{i = 1}^M \overline{\mathcal{R}}_i$. Consider an arbitrary $\mathcal{R}_i$. Recall that $V$ is piecewise linear, and by definition $V$ is affine in each $\mathcal{R}_i$. As $\mathcal{R}_i$ is an open set not containing the boundary, $V$ is continuously differentiable on $\mathcal{R}_i$. 
    By the hypothesis, $D^+_fV(\mathbf{x}) < 0$ for all $\mathbf{x} \in \mathcal{R}_i$.

    For any $\delta > 0$, define the auxiliary function
    \begin{equation}
        D^+_fV_\delta(\mathbf{x}) = \frac{V(\mathbf{x} + \delta f(\mathbf{x})) - V(\mathbf{x})}{\delta}.
    \end{equation}
    Note that $D^+_fV_\delta(\mathbf{x})$ is continuous, as both $V$ and $f$ are continuous. Consider an arbitrary $\mathbf{x}_0 \in \partial \overline{\mathcal{R}}_i$. Define a sequence $\{\mathbf{x}_k\}_{k = 1}^{\infty}$ such that $\mathbf{x}_k \to \mathbf{x}_0$ as $k \to \infty$, where $\mathbf{x}_k \in \mathcal{R}_i$ for all $k$.

    By the continuity of $D^+_fV_\delta(\cdot)$,
\begin{align}
    D^+_fV(\mathbf{x}_0) &= \limsup_{\delta \to 0^+} D^+_fV_\delta(\mathbf{x}_0) \nonumber \\
    &= \limsup_{\delta \to 0^+} \lim_{k\to\infty} D^+_fV_\delta(\mathbf{x}_k).
\end{align}
Since $V$ is affine in $\mathcal{R}_i$ and the Lie derivative is strictly negative in the interior, $D^+_fV_\delta(\mathbf{x}_k) < 0$ for any $\delta$ if $\delta$ is chosen such that $\mathbf{x}_k + \delta f(\mathbf{x})$ remains inside $\mathcal{R}_i$. 

Then, 
\begin{align}
    D^+_fV(\mathbf{x}_0) &= \limsup_{\delta \to 0^+} \lim_{k\to\infty} D^+_fV_\delta(\mathbf{x}_k) \nonumber \\
    &= \limsup_{\delta \to 0^+} b_\delta
\end{align}
where $b_\delta$ is a sequence of strictly negative real numbers whose limit supremum converges to $b\in \mathbb{R}_{\leq 0}$. Thus,
\begin{equation}
\label{eq:boundary_0}
    D^+_fV(\mathbf{x}_0) = \limsup_{h \to 0^+} b_\delta  = b\leq 0.
\end{equation}

Note that the above $b$ represents the Dini derivative at the boundary of a region. If $ D_f^+V(\mathbf{x})$ is not strictly negative at a boundary, then the trajectories on the boundary are not immediately attracted to the origin. 
Consider the pathological case where a trajectory on the boundary is not attracted to the origin. Recall that $D_f^+V(\mathbf{x}) < 0$ for all $\mathbf{x}$ in the interior of a region. Thus, a trajectory that does not converge to the origin cannot move away from the origin, or otherwise $ D_f^+V(\mathbf{x}) > 0$. Therefore, a trajectory that does not converge to the origin can only have $ D_f^+V(\mathbf{x}) = 0$, which indicates the presence of a limit cycle on a level set. A limit cycle can only occur on a level set, otherwise the Dini derivative would be nonzero. As shown in \eqref{eq:boundary_0}, $ D_f^+V(\mathbf{x}) =  0$ can only occur at the boundary of a region. In other words, a limit cycle can only occur exactly on a closed sequence of boundaries. Furthermore, the limit cycle and the associated sequence of boundaries must coincide with a level set.

 Note that since $V$ is a piecewise linear continuous function, the level sets of $V$ are also polytopes.
 However, by Assumption~\ref{assump:smooth_dynamics}, $f(\mathbf{x})$ is smooth. A limit cycle that moves exactly along a sequence of boundaries of regions on a level set would imply that $f(\mathbf{x})$ is nonsmooth, as $f(\mathbf{x})$ must follow a polyhedral path. Therefore, a limit cycle cannot exist at the boundaries on a level set, and thus $f(\mathbf{x})$ is asymptotically stable. Thus, $V$ has a non-positive Dini derivative everywhere in $\mathcal{B}$ for the system. Thus, $V$ is a weak Lyapunov function. 
\end{proof}

The above result shows that, under Assumption~\ref{assump:smooth_dynamics}, verifying Condition~\ref{cond:dec} in the interior of the linear activation regions is sufficient to guarantee asymptotic stability. We now show how the verification of the interior of each $\mathcal{R}_i$ is done in practice. 
 
Recall that $\nabla V(\mathbf{x}) \cdot f(\mathbf{x}) = D_f^+V(\mathbf{x})$ for all $\mathbf{x} \in \mathcal{R}_i$ for each $i\in [M]$, 
since $V$ is continuously differentiable in the interior of each linear activation region. Since the gradient of $V$ is constant in each $\mathcal{R}_i$, any variation in the Dini derivative is due solely to the system dynamics, so we can express
\begin{equation}\label{eq: region_dini}
    D^+_fV(\mathbf{x}) = \nabla V |_{\mathcal{R}_i} \cdot f(\mathbf{x})
\end{equation}
for $\mathbf{x} \in \mathcal{R}_i$. To guarantee that a counterexample is detected if it exists, we provide the following proposition. 
\begin{proposition}
    \label{prop: test3}
      For a region $\mathcal{R}$, there exists an $\mathbf{x}_0 \in \mathcal{B}$ such that \mbox{$\nabla V|_{\mathcal{R}} \cdot f(\mathbf{x}_0) \geq 0$} if and only if there exists $\mathbf{x} \in \mathcal{R}$ such that \mbox{$\max_{\mathbf{x} \in \mathcal{R}} \nabla V|_{\mathcal{R}} \cdot f(\mathbf{x}) \geq 0$.}
    \label{thm: optimize}
\end{proposition}

\begin{proof}
    Suppose there is an $\mathbf{x}_0 \in \mathcal{B}$ such that \mbox{$\nabla V |_{\mathcal{R}} \cdot f(\mathbf{x}_0) \geq 0$}, where $\mathbf{x}_0$ belongs to some $\mathcal{R}$. Then \mbox{$0 \leq \nabla V |_{\mathcal{R}} \cdot f(\mathbf{x}_0) \leq \max_{\mathbf{x} \in \mathcal{R}} \nabla V |_{\mathcal{R}} \cdot f(\mathbf{x})$}, completing the forward direction. 

    Now suppose that $\max_{\mathbf{x} \in \mathcal{R}}\nabla V |_{\mathcal{R}} \cdot f(\mathbf{x}) \geq 0$. Then there must exist some $\mathbf{x}_0 \in \mathcal{R}$ such that \mbox{$\max_{\mathbf{x} \in \mathcal{R}} \nabla V |_{\mathcal{R}} \cdot f(\mathbf{x}) \geq \nabla V |_{\mathcal{R}} \cdot f(\mathbf{x}_0) \geq 0$} as desired. 
\end{proof}

Proposition~\ref{thm: optimize} yields the following for verifying Condition~\ref{cond:dec}.

\begin{test}[Negative Dini Derivative — Condition~\ref{cond:dec}]
    \label{test:cond:dec}
     Consider an activation region $\mathcal{R}$. Let  $\mathbf{x}^* = \arg\max_{\mathbf{x} \in \mathcal{R} \setminus \mathbf{0}} \nabla V | _{\mathcal{R}} \cdot f(\mathbf{x})$.
    If $\nabla  V |_{\mathcal{R}} \cdot f(\mathbf{x^*}) \geq 0$, return $\mathbf{x}^*$ as a counterexample.
    Otherwise, Condition~\ref{cond:dec} is satisfied in $\mathcal{R}$.
\end{test}

We formally solve Problem~\ref{prob:dec} with the following corollary:

\begin{corollary}\label{corollary:asymp_stable}
    Suppose Assumptions~\ref{assump:roi}, ~\ref{assump:smooth_dynamics}, and ~\ref{assump:given_v} hold. Further, suppose that $V$ successfully passes Tests~\ref{test:cond:zero_origin}, ~\ref{test:cond:positivity}, and~\ref{test:cond:dec}. Then, the given system~\eqref{eq:original_system} is asymptotically stable to the origin within the domain of interest $\mathcal{B}$.
\end{corollary}
\begin{proof}
    Since $V$ passes Tests~\ref{test:cond:zero_origin} and~\ref{test:cond:positivity}, it is known that $V$ is positive-definite by Lemma~\ref{lemma: answer_prob2}. By Test~\ref{test:cond:dec}, it is verified that the Dini derivative of $V$ is strictly negative in the interior of each $\mathcal{R}_i$. Theorem~\ref{thm: interior_ver} proved that this is sufficient to show that $V$ is a weak Lyapunov function for~\eqref{eq:original_system} within $\mathcal{B}$, which proves the system is asymptotically stable to the origin.
\end{proof}

\subsection{Sound Decrease Condition Certification}\label{sec:sub:decrease}

This subsection describes our implementation of Test~\ref{test:cond:dec}.
Although Theorem~\ref{thm: np_dec_test} indicates that Problem~\ref{prob:dec} is intractable in general, certain classes of dynamical systems admit efficient verification. For example, continuous piecewise affine dynamics reduce to a linear program on each region~\cite{SamanipourTAC}, while convex or concave dynamics can be optimized efficiently using standard techniques. However, for general nonlinear dynamics, the objective is typically nonconvex, motivating the need for certified bounding techniques.

\paragraph{Hyperlyve Branch-and-Bound with LP Tightening} Hyperlyve uses the LiRPA framework~\cite{xu2020automatic} to compute a certified affine upper bound of a region's Dini derivative on a hyperrectangular space $\mathcal{S}$.
The upper bound is a hyperplane expressed as a matrix and offset,
\begin{equation}\label{eq:lirpa_box}
    D_f^+V(\mathbf{x}) =
    \nabla V|_\mathcal{R}\cdot f(\mathbf{x}) \leq \mathbf{A}_u\mathbf{x}+\mathbf{b}_u \quad \forall \mathbf{x} \in \mathcal{R} \cap \mathcal{S}.
\end{equation}
If this hyperplane is strictly negative then Test~\ref{test:cond:dec} is immediately satisfied.
However, $\mathcal{R}$ is generally not aligned with an axis (oblique), so the volume of $\mathcal{S}$ is often much larger than that of $\mathcal{R}$, causing LiRPA to produce an overly conservative bound, rendering Test~\ref{test:cond:dec} inconclusive.
In this case, Hyperlyve solves a small linear program to tighten the bound before bisecting $\mathcal{S}$ and branching, as in~\cite{yarbrough2026fitlrp}.
Branch-and-bound repeats until Test~\ref{test:cond:dec} is satisfied or a counterexample is found.

\paragraph{Origin Exclusion} Recall that by Corollary~\ref{corollary:vertex_origin}, a CPA neural Lyapunov function must have a vertex at the origin.
The Lyapunov decrease condition (Condition~\ref{cond:dec}) excludes the origin because an asymptotically stable system requires $f(\mathbf{0})=0$ by definition.
Therefore, when LiRPA bounds the Dini derivative of a region $\mathcal{R}$ that has a vertex at the origin, the resulting hyperplane must necessarily be non-negative at $\mathbf{x} = \mathbf{0}$. 

To verify the decrease condition on such regions, we must certify that the bound is strictly negative everywhere on $\mathcal{R} \cap \mathcal{S}$ except at the origin. However, floating-point numerical tolerances make excluding precisely zero infeasible. In practice, a small numerical ``hole'' around the origin must be omitted from certification~\cite{chang2019neural,zhou2022neural}. Formally, we restrict our certification domain by enforcing an $L_\infty$-ball exclusion, yielding the tightened bounding condition:
\begin{equation}\label{eq:lirpa_box_origin}
    D_f^+V(\mathbf{x}) \leq \mathbf{A}_u\mathbf{x}+\mathbf{b}_u \quad \forall \mathbf{x} \in \mathcal{R} \cap \mathcal{S} \quad \text{s.t.} \quad \|\mathbf{x}\|_\infty \geq h,
\end{equation}
where $h > 0$ defines the radius of the excluded hole.

For the experiments in Section~\ref{sec:experiments} we set $h = 10^{-6}$, approaching floating-point precision.
The radius of our excluded hole $h$ is 2-4 orders of magnitude smaller than that of competing verifiers. While approaches that use smooth Lyapunov methods must verify $V$ in conjunction with $f$, our non-smooth $V$ has a constant gradient within each region. As a result, we only need LiRPA to bound $f$, which requires fewer relaxations, resulting in tighter bounds and superior numerical stability at small values.

\subsection{Complexity of Neural Lyapunov Verification}

We previously proved that Problem~\ref{prob:pos_def} is co-NP-complete. Now suppose we are given a positive-definite $V$ and are tasked with verifying the Lyapunov decrease condition (Problem~\ref{prob:dec}).
This task is distinct from merely proving a property for a ReLU network because it is inextricably coupled with nonlinear dynamics~\eqref{eq:original_system}.

\begin{theorem}
    \label{thm: np_dec_test}
    Problem~\ref{prob:dec} is co-NP-complete. 
\end{theorem}
\begin{proof}
Let $T$ denote Problem~\ref{prob:dec}. Then by Proposition~\ref{prop: test3}, we can decompose Problem~\ref{prob:dec} such that $T = A_1 \land A_2 \land \dots A_M$, where $A_i$ represents the decision problem of whether $\nabla V |_{\mathcal{R}_i} \cdot f(\mathbf{x})$ is negative for all $\mathbf{x} \in \mathcal{R}_i$. Taking the complement, we see that $\neg T = (\neg A_1) \lor (\neg A_2) \lor \dots (\neg A_M)$, where $\neg A_i$ represents the decision problem of determining if there exists a $\mathbf{x} \in \mathcal{R}_i$ such that $\nabla V |_{\mathcal{R}_i} \cdot f(\mathbf{x})$ is nonnegative. 

Murty and Kabadi~\cite{murty1987some} proved that deciding whether there exists an $\mathbf{x} \in \mathcal{P}$ satisfying $G(\mathbf{x}) \geq 0$ is NP-complete when $G$ is a quadratic function over a polyhedral set $\mathcal{P}$. We develop a polynomial-time reduction from this problem to $\neg A_i$: determining whether there exists an $\mathbf{x}$ within a polytope $\mathcal{R}$ such that $\nabla V (\mathbf{x}) \cdot f(\mathbf{x}) \geq 0$, given a positive-definite $V$ and dynamics $f(\mathbf{x})$. This target decision problem belongs to NP because a counterexample point $\mathbf{x}$ provides a polynomial-time verifiable certificate.

    Let $G(\mathbf{x}) = g_1(\mathbf{x}) + g_2(\mathbf{x}) + \dots + g_p(\mathbf{x})$. Note that some of the $g_i(\mathbf{x})$ could be zero if $G(\mathbf{x})$ has less than $p$ terms. We then rewrite $G$ as
    $$G(\mathbf{x}) = \begin{bmatrix}
        1 & \dots & 1
    \end{bmatrix} 
    \begin{bmatrix}
        g_1(\mathbf{x}) \\
        \vdots \\
        g_p(\mathbf{x})
    \end{bmatrix}.$$
    
    Given the region $\mathcal{R}_i = \text{Int}(\mathcal{P})$, let $\mathbf{W}$ and $\mathbf{b}$ be a matrix and vector, respectively such that $\mathcal{R}_i = \{\mathbf{x} \in \mathbb{R}^p: \mathbf{W} \mathbf{x} < \mathbf{b}\}$. Now define the ReLU neural network $V$ with $\mathbf{W}$ and $\mathbf{b}$ and the corresponding weights and biases such that $\nabla V |_{\mathcal{R}_i} = \begin{bmatrix}
         1 & \dots & 1
    \end{bmatrix}$. . Further, define the dynamics $f(\mathbf{x}) = \begin{bmatrix}
         g_1(\mathbf{x}) & \dots & g_p(\mathbf{x})
     \end{bmatrix}^T $. 
Thus, we have successfully constructed an instance of Problem~\ref{prob:dec}, namely, $\neg A_i$. Then an $\mathbf{x}$ satisfies \mbox{$\nabla V(\mathbf{x}) \cdot f(\mathbf{x}) \geq 0$} if $G(\mathbf{x}) \geq 0$ is satisfied. Thus, $\neg A_i$ is NP-complete. Since $\neg T$ is the finite union of NP-complete problems, $\neg T$ must also be NP-complete. Therefore, $T$ is co-NP-complete as desired.
\end{proof}

We have shown that Tests~\ref{test:cond:zero_origin}, ~\ref{test:cond:positivity}, and~\ref{test:cond:dec} are sufficient to prove whether $V$ is a neural Lyapunov function for a given smooth dynamical system. 
However, these Tests require full knowledge of each polyhedral linear activation region. We now introduce our methodology to discover these regions. 

\section{The Hyperlyve Framework}
\label{section:region_enum}

This section presents our complete verification framework. We first introduce a parallel recursive algorithm for enumerating activation regions in deep ReLU networks (Problem~\ref{prob:enumerate}). We then integrate this enumerator into the complete Hyperlyve verifier (Algorithms for Problems~\ref{prob:pos_def} and~\ref{prob:dec}) and establish its formal soundness and completeness for Problem~\ref{problem:main_idea}.

\subsection{Exact Region Enumeration}
The regional stability certificates derived in Sections~\ref{sec:prove_pos_def} and~\ref{section:verifying_lyap} reduce the continuous verification problem to certifying conditions over individual convex polytopes.
Unfortunately, merely counting the affine regions of even a single-hidden-layer ReLU network is \#P-complete~\cite{stargalla2025computational}, a class of counting problems at least as hard as NP-hard problems~\cite{valiant1979complexity}. Despite this computational hardness, exact region enumeration remains practical and effective: decomposing the state space into individual affine polytopes enables parallel optimization that significantly outperforms global reasoning over non-smooth neural networks (as demonstrated empirically in Section~\ref{sec:experiments}).

\paragraph{Hierarchical Tree Structure of Regions}
Deep feedforward ReLU networks partition the input space into convex polyhedral regions, each associated with an affine local mapping~\cite{raghu2017expressive}. The polyhedral structure arises because each layer applies a recursive composition of affine transformations and piecewise-linear activations~\cite{montufar2014number}, yielding effective regional weights $\mathbf{W}_e$ and biases $\mathbf{b}_e$.

The layer-wise recursive composition of activation regions defines a hierarchical tree in which every child region in the current layer has only a single parent region in the previous layer; furthermore, the children from any region in the current layer will not intersect the children from a different region in the current layer:
\begin{itemize}
    \item Each region $\mathcal{R}$ in layer $\ell$ has a unique parent region $\mathcal{P}$ in layer $\ell-1$.
    \item Each parent forms the root of an independent subtree, meaning child regions within a subtree do not depend on adjacent cells belonging to other parents.
    \item The leaves at the final hidden layer form a polyhedral complex tiling the domain of interest $\mathcal{B}$.
\end{itemize}
Hyperlyve exploits this structural independence to process subtrees concurrently in parallel worker threads without requiring shared locking or inter-process communication, avoiding the expensive network-flattening steps used in prior work~\cite{vincent2025tnnls, gaines2026characterizing}.

\paragraph{Facet Crossing and Enumeration}
The facets of a region $\mathcal{R}$ are determined by the non-redundant linear constraints of the parent region $\mathcal{P}$ in layer $\ell-1$ and the arrangement $\mathcal{A}^e_{\ell}$, which denotes the arrangement formed by effective parameters at layer $\ell$:
\begin{equation}\label{eq:facets}
\begin{split}
    \text{facets}(\mathcal{R}) = \{ H \cap \overline{\mathcal{R}} \mid \;& H \in \text{facets}(\mathcal{P}) \cup \mathcal{A}^e_{\ell}, \\
    & \dim(H \cap \overline{\mathcal{R}}) = p-1 \}.
\end{split}
\end{equation}

\begin{algorithm}[t]
\caption{\textsc{Facet BFS}}
\label{alg:layer_bfs}
\begin{algorithmic}[1]
\Require ReLU network $V$, parent region $\mathcal{P}$, layer $\ell$.
\Ensure Iterate each layer-$\ell$ region of $V$ within $\mathcal{P}$.

\State $\mathbf{W}_e, \mathbf{b}_e \gets \textsc{EffectiveWeights}(V, \mathcal{P})$
\State $\mathcal{R}_0 \gets \Call{Seed}{\mathcal{P}, \mathbf{W}_e, \mathbf{b}_e}$
\State \Yield{$\mathcal{R}_0$}\label{alg:layer_bfs:yield1} \Comment{Yield initial region for streaming recursion.}
\State $\textsc{Visited} \gets \{\sigma(\mathcal{R}_0)\}$
       \Comment{Boolean activation pattern.}
\State $\mathcal{Q} \gets \emptyset$;\quad
        $\Call{Enqueue}{\mathcal{Q}, \mathcal{R}_0}$

\While{$\mathcal{Q} \neq \emptyset$}
    \State $\mathcal{R} \gets \Call{Dequeue}{\mathcal{Q}}$;\quad
    \For{$H \in \text{facets}(\mathcal{R)}$, $H \notin \text{facets}(\mathcal{P})$} \Comment{Eq.~\eqref{eq:facets}}
        \LComment{Flip the facet, including collinear hyperplanes.}
        \State $\mathbf{s}' \gets \sigma(\mathcal{R})$ with $s'_i = -s_i \;\; \forall i : H_i = H$
        \If{$\mathbf{s}' \in \textsc{Visited}$}
            \State \textbf{continue}
        \EndIf
        \State $\textsc{Visited} \gets \textsc{Visited} \cup \{\mathbf{s}'\}$
        \State $(A', \mathbf{b}') \gets \text{facets}(\mathcal{R})$ with $H$ constraints negated
            \label{alg:layer_bfs:flip}
        \State $\mathbf{c}, r \gets \Call{ChebyshevCenter}{A', \mathbf{b}'}$\label{alg:layer_bfs:chebyshev}
        \If{$r \le \epsilon$}
            \State \textbf{continue} \Comment{No full-dimensional cell}
        \EndIf
        \State $\mathcal{R}' \gets \Call{QHull}{A', \mathbf{b}', \mathbf{c}}$\label{alg:layer_bfs:qhull}\Comment{Facets \& vertices\cite{barber1996quickhull}}
        \State $\Call{Enqueue}{\mathcal{Q}, \mathcal{R}'}$;\quad \Yield{$\mathcal{R'}$}\label{alg:layer_bfs:yield2}
    \EndFor
\EndWhile
\end{algorithmic}
\end{algorithm}

\begin{algorithm}[ht]
\caption{Hyperlyve: Neural Lyapunov Verification}
\label{alg:main_verify}
\begin{algorithmic}[1]

\Require ReLU network $V$, dynamics $f$, compact domain $\mathcal{B}$.
\Ensure Counterexample set \textsc{Cex}; $\textsc{Cex} = \emptyset \iff$ verified.

\State $\textsc{Cex} \gets \emptyset$
\If{$V(\mathbf{0}) \neq 0$} $\textsc{Cex} \gets \{\mathbf{0}\}$ \Comment{Test \ref{test:cond:zero_origin}: Zero origin}
\EndIf
\LComment{Stream BFS regions to the parallel worker pool.}
\ForAll{$\mathcal{R} \in \Call{FacetBFS}{V, \mathcal{B}, \ell=1}$}
    \State $\textsc{Cex} \gets \textsc{Cex} \cup \Call{ProcessRegion}{\mathcal{R}, \ell=1}$
\EndFor
\State \Return \textsc{Cex}

\Statex

\Function{ProcessR}{$\mathcal{R}, \ell$}
    \If{$\ell = \Call{LastLayer}{V}$}
        \LComment{Test~\ref{test:cond:positivity}: Positive-definiteness}
        \State $\mathcal{X}_{\text{pos}} \gets \{\mathbf{x}^* \in \operatorname{vert}(\overline{\mathcal{R}}) \setminus \{\mathbf{0}\} : V(\mathbf{x}^*) \leq 0\}$

        \LComment{Test~\ref{test:cond:dec}: Decrease condition with~\eqref{eq:lirpa_box_origin} and BaB}
        \State $\mathcal{X}_{\text{dec}} \gets \{\mathbf{x}^* \in \mathcal{R} \setminus \{\mathbf{0}\} \mid \newline
        \mathbf{x}^* \in \arg \max \nabla V|_\mathcal{R} \cdot f(\mathbf{x}), \; ||\mathbf{x}^*||_\infty \geq h, \newline
        \nabla V|_\mathcal{R} \cdot f(\mathbf{x}^*) \geq 0\}$\label{alg:main_verify:dec_cex}
        
        \State \Return $\mathcal{X}_{\text{dec}} \cup \mathcal{X}_{\text{pos}}$
    \Else
        \LComment{Process stream of yielded region in parallel.}
        \State $\mathcal{C} \gets \Call{FacetBFS}{V, \mathcal{R}, \ell+1}$ \Comment{Alg.~\ref{alg:layer_bfs}}
        \State \Return $\displaystyle\bigcup_{\mathcal{R}' \in \mathcal{C}} \Call{ProcessR}{\mathcal{R}', \ell+1}$ \label{alg:main:worker_share}
    \EndIf
\EndFunction

\end{algorithmic}
\end{algorithm}

Algorithm~\ref{alg:layer_bfs} shows the procedure for facet breadth-first search (BFS). The local visited set is tracked by the activation pattern $\sigma(\cdot)$, a Boolean vector of active/inactive neurons, which is injective to a specific region.
After crossing a facet from the current region into an unvisited neighboring region, we check feasibility of the new region via its Chebyshev center (ball of largest inscribed radius) by solving a small LP that is warm-started from the previous region's basis. We identify the facets of each feasible region in line~\ref{alg:layer_bfs:qhull} using QHull~\cite{barber1996quickhull} to solve the dual convex hull, which has the additional benefit of yielding vertices.
QHull is strictly better than naive per-neuron checks when $p<N_\ell$, which is nearly the ubiquitous case for dynamical systems. As we will discuss in Section~\ref{sec:sub:complex}, QHull enjoys superior theoretical complexity compared to pruned LP approaches in low-to-moderate state dimensions.

The \textbf{yield} calls on lines~\ref{alg:layer_bfs:yield1} and~\ref{alg:layer_bfs:yield2} stream feasible regions to the parallel worker pool. Idle workers can process a region yielded from any subtree, so new depth-first search trees are spawned while local BFS is ongoing.
The next subsection describes how this traversal feeds directly into Algorithm~\ref{alg:main_verify}, enabling Hyperlyve to certify the entire Lyapunov candidate in parallel, without global coordination bottlenecks

\subsection{Hyperlyve Algorithm}

Hyperlyve performs exact region enumeration and Lyapunov verification of deep neural networks via parallel recursive facet-march. We formally present Hyperlyve in Algorithm~\ref{alg:main_verify}:
\begin{itemize}
    \item \textbf{Bounded Breadth-First Search:} Conduct a BFS facet march to discover the regions formed by  the current layer, $\ell$, within given compact domain bounds (Algorithm~\ref{alg:layer_bfs}).
    \item \textbf{Task Dispatch:} Yield each discovered region as an independent parent task to a parallel worker pool.
    \item \textbf{Depth-First Recursion:} Each worker conducts a BFS in layer $\ell+1$, with facets of the layer-$\ell$ parent region as the domain bounds. Each BFS emits recrusive tasks.
    \item \textbf{Lyapunov Verification:} Regions produced from the final hidden layer $L$ are eagerly certified against the weak Lyapunov conditions, optionally leading to an early exit if a counterexample is found.
\end{itemize}

Each Facet BFS (Algorithm~\ref{alg:layer_bfs}) is only ever handled by a single worker. Yet, since each parent task seeds a disjoint subtree, no two workers ever visit the same region, which eliminates the need for a global visited. Meanwhile, tasks in the next layer of the network may be freely distributed to idle workers (Algorithm~\ref{alg:main_verify}, line~\ref{alg:main:worker_share}).
We seed the worker pool with the first layer of $V$ by Sobol sampling $\mathcal{B}$~\cite{owen1998scrambling}.

To solve the nonconvex global optimization in line~\ref{alg:main_verify:dec_cex}, we first apply a cheap pre-filter: local SLSQP optimization~\cite{kraft1988software} of the Dini derivative seeded from the region's Chebyshev center produces a counterexample, return the counterexample.
Otherwise, we begin a branch-and-bound (BaB) procedure (Section~\ref{sec:sub:decrease}) that uses LiRPA~\cite{xu2020automatic} to produce affine upper bounds, which we then tighten against the constraints in~\eqref{eq:lirpa_box_origin} using a series of linear programs, as in~\cite{yarbrough2026fitlrp}.

Although elements of our algorithm appear in isolation across prior literature---such as bounded recursive traversal~\cite{xu2021traversing} or facet-crossing rather than neuron-crossing~\cite{vincent2025tnnls}---Hyperlyve combines these techniques and realizes maximum parallelization with minimal overhead.
Note that verification can be gated and Hyperlyve can return the $H$-representation of a network absent of dynamics.

\subsection{Hyperlyve Guarantees}

We now show that Hyperlyve is a sound and complete solution.

\begin{lemma}
    \label{lemma: alg_works}
    Given a ReLU neural network $V$ and compact domain $\mathcal{B}$, Algorithm~\ref{alg:layer_bfs} will return every linear activation  within $\mathcal{B}$. 
\end{lemma}
\begin{proof}
    We proceed by contradiction. Suppose Algorithm~\ref{alg:layer_bfs} returns a set of regions $\mathcal{T} = \{\mathcal{R}_1, \dots, \mathcal{R}_t\}$, for some $t \in \mathbb{N}$. Further, suppose that $\mathcal{R}_0$ is also a linear activation region and $\mathcal{R}_0 \notin \mathcal{T}$. Let $\mathcal{S} = \{ \text{facets}(\mathcal{R}_1), \dots, \text{facets}(\mathcal{R}_t)\}$ be the set of facets for the discovered regions. Let $F(i, \text{facets}(\mathcal{R}_j))$ denote a function that flips the $i$th facet of the $j$th region. 

    Since $\text{facets}(\mathcal{R}_0) \not\subset \mathcal{S}$, then $F(i, \text{facets}(\mathcal{R}_0)) \not\subset \mathcal{S}$ for any $i$. If there exists an $i$ such that $F(i, \text{facets}(\mathcal{R}_0)) \subset \mathcal{S}$, then Algorithm~\ref{alg:layer_bfs} would have discovered this region, as it discovers regions by flipping each facet of the discovered regions. $F(i, \text{facets}(\mathcal{R}_0)) \not\subset \mathcal{S}$ implies that $\mathcal{R}_0$ is not adjacent to any region in $\mathcal{T}$, and thus $\overline{\mathcal{R}}_0$ is disjoint from the closure of the discovered activation regions. Such a disjoint activation region cannot exist as $V$ is piecewise linear and defined everywhere in the state space. Therefore, every activation region is discovered by Algorithm~\ref{alg:layer_bfs}, completing the proof. 
\end{proof}

\begin{theorem}
\label{thm:sound_complete}
Suppose Assumptions~\ref{assump:roi}, ~\ref{assump:smooth_dynamics}, and~\ref{assump:given_v} hold.
    Suppose the maximum iterations parameter for branch-and-bound of the Dini derivative is $I_{\text{max}} = \infty$.
    Then, Algorithm~\ref{alg:main_verify} is sound and complete.
\end{theorem}
\begin{proof}
    Suppose ReLU neural network $V$, domain of interest $\mathcal{B}$, and dynamics $f$ from the assumptions are passed into Hyperlyve (Algorithm~\ref{alg:main_verify}) and Hyperlyve returns that $V$ is a valid Lyapunov function for $f$ over $\mathcal{B}$.  

    Let $\mathcal{T} = \{\mathcal{R}_1, \dots, \mathcal{R}_M\}$ denote the set of linear activation regions found by Algorithm~\ref{alg:layer_bfs}. Appealing to Lemma~\ref{lemma: alg_works}, it is known that $\mathcal{T}$ is not only the set of discovered activation regions, but also the complete set of all activation regions induced by $V$. Hyperlyve returning true indicates that Tests~\ref{test:cond:zero_origin}, \ref{test:cond:positivity}, and~\ref{test:cond:dec} are successful for each activation region found. Appealing to Corollary~\ref{corollary:asymp_stable}, $V$ must be a weak Lyapunov function for $f$ over $\mathcal{B}$. Therefore, Algorithm~\ref{alg:main_verify} is sound. 

    Suppose $V$ is a weak Lyapuonv function for the dynamics $f$ over a domain of interest $\mathcal{B}$. Then, by Definition~\ref{def:weak_lyap}, $V(\mathbf{0}) = 0$, and Test~\ref{test:cond:zero_origin} is passed successfully. Again, by Definition~\ref{def:weak_lyap}, $V$ is strictly positive everywhere in the domain of interest excluding the origin. Then $V$ is positive definite over every linear activation region (except where it is zero at the vertex that lies on the origin by Corollary~\ref{corollary:vertex_origin}), and thus each region passes Test~\ref{test:cond:positivity} successfully. Applying Definition~\ref{def:weak_lyap} again, $V$ must decrease along all trajectories. Then the Dini derivative at the interior of each activation must be strictly negative.
    Thus, since $I_{\max} = \infty$, branch-and-bound from~\cite{yarbrough2026fitlrp} will eventually prove that each region successfully passes Test~\ref{test:cond:dec}.
    
    We have now shown that $V$ being a valid Lyapunov function implies that each activation region passes Tests~\ref{test:cond:zero_origin}, \ref{test:cond:positivity}, and \ref{test:cond:dec}. By Lemma~\ref{lemma: alg_works}, Hyperlyve has access to the exact set of linear activation regions, and will proceed to perform Tests~\ref{test:cond:zero_origin}, \ref{test:cond:positivity}, and \ref{test:cond:dec} on each activation region. Each test will be successfully passed as shown above, and Algorithm~\ref{alg:main_verify} will return true. Therefore, Hyperlyve is complete. 
\end{proof}

\subsection{Algorithmic Complexity}\label{sec:sub:complex}

Hyperlyve addresses co-NP-complete decision problems (Theorems~\ref{thm: pos_np} and~\ref{thm: np_dec_test}) and thus exhibits exponential time complexity in the worst case. However, its practical algorithmic performance on average is substantially better than this theoretical bound.

\subsubsection{Worst-Case Complexity}

The computational complexity of Hyperlyve depends on the total number of polyhedral regions and the construction cost per region.
Recall that there are $M$ open linear activation regions contained in $\mathcal{B}$. The problem of bounding the number of activation regions, or equivalently measuring the expressivity of a ReLU neural network, is a well-studied problem. Several valid upper bounds for $M$ can be found in~\cite{li2025achieving, serra2018bounding, serra2020empirical, raghu2017expressive, montufar2014number}.

The Quickhull halfspace intersection algorithm is output-sensitive in complexity, but when balance conditions hold, its cost is dominated by the total number of facets it constructs, which, by McMullen's Upper Bound Theorem~\cite{mcmullen1970maximum} is itself bounded by the number of $m$ hyperplanes, $\mathcal{O}(m^{\lfloor p/2 \rfloor})$.
Since finding the Chebyshev center (Algorithm~\ref{alg:layer_bfs}, line~\ref{alg:layer_bfs:chebyshev}) requires only solving a small linear program, we assume that Quickhull step dominates the computational cost of region construction.
In the worst case, where every region is created by the first layer, all $M$ regions must be solved in each of the $L$ layers.

For verification, let $T_\text{verify}$ denote the combined cost of verifying a region under Tests~\ref{test:cond:positivity} and~\ref{test:cond:dec}. Test~\ref{test:cond:positivity} scales polynomially with the number of vertices. Test~\ref{test:cond:dec}, however, lacks a finite worst-case termination guarantee due to the input-dependent nature of branch-and-bound routines. The upper complexity bound for Hyperlyve is thus $\mathcal{O}(LM m^{\lfloor p/2 \rfloor} + MT_\text{verify})$.

\subsubsection{Average-Case Complexity} 
In practice, hyperplanes in a polyhedral complex are localized, meaning individual constraints bound only a small subset of regions. Specifically, the average number of facets per cell in a $p$-dimensional arrangement is bounded by $2p$~\cite{roudneff1991cells, gaines2026characterizing}.

Hanin and Rolnick~\cite{hanin2019deep} show that under random network assumptions the expected number of activation regions scales as $\mathcal{O}(m^p/p!)$, which is much smaller than $M$. Discounting verification cost, replacing the global number of hyperplanes, $m$, with the local expected number of facets, $2p$, reduces the expected enumeration complexity to
\begin{equation}\label{eq: expected_work}
    \mathcal{O} = \max\left(\dfrac{Lm^p}{p!},\; T_{\text{QH}}\right),
\end{equation}
where $T_{\text{QH}}$ is the time complexity of QuickHull.
For low dimensions, $p\leq3$, Quickhull is bound logarithmically~\cite{barber1996quickhull},
which is strictly better than the $p\geq4$ bound given in~\eqref{eq: expected_work}.
For low-to-moderate input dimensions and sufficiently wide networks, identifying facets with Quickhull is computationally superior to solving LPs over all $m$ network neurons.
As $p$ grows, migrating to LP-based facet enumeration coupled with a bounding box heuristic~\cite{suard2004efficient, vincent2025tnnls} likely becomes better.

\section{Training a Neural Lyapunov Function}\label{section:training}

\begin{figure}[t]
    \centering
    \includegraphics[width=\linewidth]{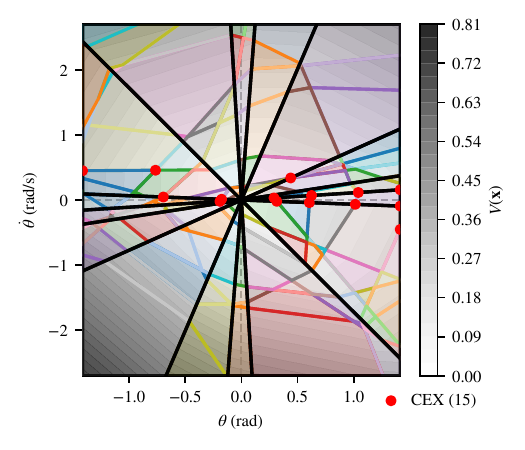}
    \caption{Hyperlyve enumeration of a candidate Lyapunov network $V$ and identification of counterexamples for an inverted pendulum with a neural network controller. $V$ has 3 hidden layers with $[8, 12, 12]$ neurons. Black lines denote first-layer ReLU activation hyperplanes, while the shaded regions and multicolored lines represent activation regions and hyperplanes of subsequent layers (Algorithm~\ref{alg:layer_bfs}). Red dots indicate the 15 counterexamples (\textsc{Cex}) identified by Algorithm~\ref{alg:main_verify}.}
    \label{fig:deep_regions}
\end{figure}

Integrating Hyperlyve as the sound and complete verifier for a CEGIS loop to learn a CPA neural Lyapunov function (recall Figure~\ref{fig:cegis_loop}) provides three key advantages compared to competing approaches: 
1) it produces a representative set of counterexamples across the domain of interest while preventing false-positive counterexamples from corrupting the training set, 
2) it yields an exact stopping criterion as soon as a weak neural Lyapunov function is synthesized,
3) Hyperlyve is faster than competing verifiers, which allows for more rapid CEGIS iterations.
This contrasts with approaches that only provide a single counterexample per iteration~\cite{chang2019neural, dai2020counter,  zhou2022neural}, rely on incomplete verifiers~\cite{richards2018lyapunov}, or use heuristics to trigger expensive verification phases~\cite{wu2023neural,yang2024lyapunov,li2025two}.
Hyperlyve's counterexample discovery is shown in Figure~\ref{fig:deep_regions}, which illustrates verification during a CEGIS run to train a three-layer Lyapunov network $V$ for an inverted pendulum with a neural network controller. Hyperlyve enumerates $V$ and performs Tests~\ref{test:cond:zero_origin}-\ref{test:cond:dec}, producing fifteen counterexamples across the domain.

\paragraph{Training Dataset} We generate the initial training dataset, $X \subset \mathcal{B}$, by Sobol sampling~\cite{owen1998scrambling} the domain of interest (note, this is a different sampling than that used to seed Hyperlyve verification). Each subsequent CEGIS iteration first receives a set of counterexamples from the Lyapunov verification procedure. We then seek to maximize the training signal that each counterexample provides to the loss function by finding the local worst-case violation.
Counterexamples to the positive-definite condition are already worst-case per region. Meanwhile, decrease-condition counterexamples are merely the first point in a region where the Dini derivative was identified as positive. Therefore, we perform an optimization round using SLSQP~\cite{kraft1988software} to find the local minimum of the Dini derivative for each counterexample (this follows line~\ref{alg:main_verify:dec_cex} in Algorithm~\ref{alg:main_verify}). Counterexamples are accumulated across iterations. We augment with a fresh Sobol sample each training iteration to increase stochasticity.

\paragraph{Loss Function}
We use AdamW to optimize our adaption of the empirical Lyapunov loss from~\cite{chang2019neural},
\begin{equation}\label{eq:loss}
 \begin{split}
   \textsc{Loss}(V, X) = {} & \alpha_1 V(\mathbf{0})^2 + 
         \frac{\alpha_2}{N} \sum_{i=1}^N \phi_\beta\bigl(-V(\mathbf{x}_i)\bigr) + \\
        & \frac{\alpha_3}{N} \sum_{i=1}^N \phi_\beta\bigl(\nabla V(\mathbf{x}_i)\cdot f(\mathbf{x}_i)\bigr),
\end{split}
\end{equation}
where we replace the standard ReLU hinge with a smooth \emph{softplus} activation:
\begin{equation}\label{eq:softplus}
    \phi_\beta(x) = \beta^{-1}\log(1 + e^{\beta x}).
\end{equation}
Here $\beta$ controls the tightness of the smooth approximation (we use $\beta=8$).
The smooth gradients of the softplus function provide a small loss for those sample points that barely satisfy the Lyapunov conditions, with the intent of pushing regions of the network near the sample to a better value.
Each term of~\eqref{eq:loss} penalize violations of a Lyapunov condition, with hyperparameter $\alpha$ tuning their relative weights (we set $\alpha_1=\alpha_2=1$ and $\alpha_3 = 1.5$ as defaults).

\paragraph{Architectural Origin Enforcement}
Lyapunov Condition~\ref{cond:zero_origin} requires $V(\mathbf{0}) = 0$, and Corollary~\ref{corollary:vertex_origin} showed that the origin must lie exactly on the vertex of an activation region.
Rather than relying on gradient descent to learn these strict constraints, we enforce them architecturally. 
To guarantee that the origin is a vertex, we set the bias vector of the first hidden layer to zero ($b_1 = \mathbf{0}$) and freeze $b_1$ throughout training. 
This forces the first-layer hyperplanes into a central arrangement intersecting at $\mathbf{0}$. 
This central arrangement is potentially less expressive than a generic arrangement, yet it is ideal for Lyapunov synthesis: it anchors the central critical point to the origin, makes the first layer positive homogeneous, and allows the deeper layers to fit complex dynamics.
In experimentation we observed that the central arrangement approach converged more quickly and reliably than when we attempted to keep the network generic by only enforcing zero-bias for $p$ first-layer neurons.

Finally, to guarantee $V(\mathbf{0}) = 0$ exactly, we update the bias of the output layer after each training iteration via:
\begin{equation}
    \text{bias}_{\text{final}} \gets \text{bias}_{\text{final}} - V_\theta(\mathbf{0}).
\end{equation}
Due to the $V(\mathbf{0})^2$ term in the loss function, this shift is typically small.

\section{Experiments}\label{sec:experiments}

We now empirically substantiate our claims:
\begin{enumerate}
    \item Hyperlyve enumerates the polyhedral complex of ReLU networks (Problem~\ref{prob:enumerate}) faster than state-of-the-art (SOTA).
    \item Hyperlyve certifies the asymptotic stability of dynamical systems (Problem~\ref{problem:main_idea}) faster, with larger input dimensions, and with a tighter origin tolerance than SOTA.
\end{enumerate}

We conducted experiments on an AMD EPYC 7763 Milan 64-Core Processor with up to 128GB of RAM.
Hyperlyve does not use a GPU, but an enterprise-grade GPU (AMD M120) was provided to tools that leverage $\alpha,\beta$-CROWN.
We set a 1-hour timeout, indicated by \TO, for each individual run; tool errors are denoted by \ERR.

\subsection{Enumeration Benchmarks}

Table~\ref{tab:enumerate} compares Hyperlyve's enumeration of the complete $H$-representation of a network in a domain to RPM~\cite{vincent2025tnnls} and Relucent~\cite{gaines2026characterizing}. To the best of our knowledge, these are the only other tools that return the exact $H$-representation required for Lyapunov verification Tests~\ref{test:cond:positivity} and~\ref{test:cond:dec}. To demonstrate Hyperlyve's scalability beyond standard stability experiments, the benchmarks include larger networks, such as Rejoin RL, where Hyperlyve parallelizes the enumeration of 6.3-million 8-dimensional regions in under 16 minutes while RPM and Relucent hit the 1-hour timeout.

\begin{table}[t]
\renewcommand{\arraystretch}{1.3}
\caption{Hyperlyve computes the $H$-representation of ReLU-induced polyhedral complexes over $150\times$ faster than competing tools. Hyperlyve was the only tool to enumerate the largest networks before the 1-hour timeout.}
\label{tab:enumerate}
\centering
\setlength{\tabcolsep}{3.5pt}
\resizebox{\columnwidth}{!}{%
\begin{threeparttable}
\begin{tabular}{@{} l c l r r r r @{}}
\toprule
& & & & \multicolumn{3}{c}{Average Execution Time (s)} \\ 
\cmidrule(l){5-7}
Network & $p$ & Neurons & Regions & \begin{tabular}[c]{@{}c@{}}RPM~\cite{vincent2025tnnls}\\ (TNNLS '25)\end{tabular} & \begin{tabular}[c]{@{}c@{}}Relucent~\cite{gaines2026characterizing}\\ (ICLR '26)\end{tabular} & \textbf{Hyperlyve} \\ 
\midrule
L-Index~\cite{he2022characterizing} & 1 & $5 \times 128$ & 236 & \textbf{2.7} & 105.2 & 4.6 \\
TaxiNet~\cite{katz2022verification} & 2 & $7\times \tnote{$\dagger$}$ & 280,217 & \TO & \TO & \textbf{77.6} \\
Lorenz Lyapunov & 3 & $2 \times 24$ & 8,222 & \textcolor{red!75!black}{\ERR} & 19.8 & \textbf{4.8} \\
Stanford Bunny~\cite{berzins2023edge} & 3 & $3 \times 16$ & 31,168 & 97.2 & 71.5 & \textbf{5.2} \\
ACAS Xu~\cite{katz2017reluplex} & 5 & $6 \times 50$ & 39,835 & 350.2 & 1212.3 & \textbf{8.0} \\
SIS6 Lyapunov & 6 & $3 \times 8$ & 284 & \textcolor{red!75!black}{\ERR} & \textbf{3.0} & 4.6 \\
LunarLander RL~\cite{brockman2016openai} & 8 & $2 \times 64$ & 734 & \textbf{3.8} & 36.5 & 4.6 \\
Rejoin RL~\cite{ravaioli2022safe} & 8 & $2 \times 256$ & 6,370,164 & \TO & \TO & \textbf{940.2} \\ 
\bottomrule
\end{tabular}
\begin{tablenotes}\footnotesize
    \item[] Networks are from VNN-COMP~\cite{brix2023vnncomp_overview} and Hyperlyve Lyapunov functions (see Table~\ref{tab:CEGIS_table}). Times reflect warm-start execution, omitting process launch and precompilation overheads for each tool. Bold values indicate the fastest time per row.
    \item[$\dagger$] TaxiNet is structured as 7 layers: $[4\times 256,\ 2\times 16,\ 1\times 8]$.
    \item[\TO] Denotes a 1-hour timeout limit.
    \item[\ERR] RPM returns an incorrect result for non-generic networks, due to many hyperplanes intersecting at the origin.
\end{tablenotes}
\end{threeparttable}%
}
\end{table}

\subsection{Asymptotic Stability Benchmarks}

\subsubsection{Dynamical Systems}

We used the following nonlinear dynamical systems for our stability experiments. These systems come from a variety of fields, illustrating the robustness of Hyperlyve.

\noindent\textbf{AKP Dynamics~\cite{ahmadi2011globally}:}
    \begin{align*}
    \dot{x}_1 &= -x_1 + x_1 x_2 \\
    \dot{x}_2 &= -x_2,
\end{align*}
which was introduced by Ahmadi, Krstic, and Parrilo, who proved that the system has no polynomial global Lyapunov function. We run \textbf{AKP-10} with $\mathcal{B}=[-10,10]^2$ and \textbf{AKP-100} $\mathcal{B}=([-100, 100] \times [-10,10])$ to highlight the consequence of different domain sizes.

\noindent\textbf{Lorenz System~\cite{lorenz1963deterministic}: }
    \begin{align*}
    \dot{x} &= \sigma(y - x) \\
    \dot{y} &= \rho x - y - xz \\
    \dot{z} &= xy - \beta z,
    \end{align*}
where $\rho = \frac{1}{2}, \beta = \frac{8}{3}$, and $\sigma = 10$.

\noindent \textbf{Networked SIS Epidemic Model~\cite{lajmanovich1976deterministic}:}
\begin{align*}
    \dot{\mathbf{x}} = -\mathbf{D}\mathbf{x} + (\mathbf{I} - \operatorname{diag}(\mathbf{x})) \mathbf{B}\mathbf{A}\mathbf{x}
\end{align*}
where  $x \in [0,1]^p$ represents the infection states across $p$ nodes, and $\mathbf{A}, \mathbf{B}, \mathbf{D} \in \mathbb{R}^{p \times p}$ define the network characteristics. Here, $\mathbf{A}$ is the adjacency matrix generated by an Erdős-Rényi random graph~\cite{erdos1959random} ($\text{seed}=37$), while $\mathbf{B}$ and $\mathbf{D}$ are diagonal matrices containing infection and recovery rates, respectively.

\subsubsection{Controlled Systems} For the following systems, a neural network provides control input $u$ is provided by a neural network controller.
Each controller has $\tanh$ activations and hidden size expressed as layers-of-neurons: $L \times N$.

\noindent\textbf{Inverted Pendulum (IPNN): }
\begin{align*}
    \begin{bmatrix}
        \dot{\theta}_1 \\
        \dot{\theta}_2
    \end{bmatrix} = 
    \begin{bmatrix}
        \theta_2 \\
       -\frac{\beta}{m l^2} \theta_2 + \frac{g}{l} \sin (\theta_1) + \frac{1}{ml^2}u
    \end{bmatrix},
\end{align*}
where $\beta = 1.0, m = 5.0, l = 5.0, g = 9.81$; controller is $2 \times 6$.

\noindent\textbf{Two-Link Planar Robot Arm (TLPRA)~\cite{sun2021learning}:}
\begin{align*}
    \begin{bmatrix}
        \dot{q} \\
        \ddot{q}
    \end{bmatrix} =
    \begin{bmatrix}
        \dot{q} \\
        M(q)^{-1}\bigl(u - c(q,\dot{q}) - G(q)\bigr)
    \end{bmatrix},
\end{align*}
with joint angles $q = [q_1, q_2]^\top$, torques $u = [u_1, u_2]^\top$, and
\begin{align*}
    M(q) &=
    \begin{bmatrix}
        \alpha + \beta + 2\eta\cos q_2 & \beta + \eta\cos q_2 \\
        \beta + \eta\cos q_2 & \beta
    \end{bmatrix}, \\
    c(q,\dot{q}) &=
    \begin{bmatrix}
        -\eta\,(2\dot{q}_1\dot{q}_2 + \dot{q}_2^2)\sin q_2 \\
        \eta\,\dot{q}_1^2\sin q_2
    \end{bmatrix}, \\
    G(q) &=
    \begin{bmatrix}
        \alpha g\cos q_1 + \eta g\cos(q_1 + q_2) \\
        \eta g\cos(q_1 + q_2)
    \end{bmatrix},
\end{align*}
where $\alpha = 3.1$, $\beta = 2.3$, $\eta = 2.3$, and $g = 9.8$ (link masses $m_1 = 0.8$, $m_2 = 2.3$ and unit link lengths). Equilibrium $q = \dot{q} = 0$ is held by the constant gravity-compensating torque $u^\ast = [(\alpha+\eta)g,\ \eta g]^\top$;
$\mathcal{B} = [-\frac{\pi}{8}, \frac{\pi}{8}]^4$; The controller is $4 \times 64$ and was trained with Sun et al.'s Contraction Metric~\cite{sun2021learning}.

\input{lyapunov_table}

\subsubsection{Competing Systems}
Our Hyperlyve CEGIS framework (Section~\ref{section:training}) outperforms existing approaches for proving the asymptotic stability of nonlinear continuous-time systems (Problem~\ref{problem:main_idea}).
We acknowledge that some of these verifiers are capable of synthesizing controllers, which Hyperlyve does not yet support; thus, we restricted our experiments to verifying pre-trained controllers.
Additionally, while some systems certify a region of attraction (ROA), Hyperlyve certifies a given polyhedral domain $\mathcal{B}$, typically resulting in a significantly larger certified volume.
For the decrease condition, Hyperlyve excludes a tiny region around the origin where $||\mathbf{x}||_\infty \leq h$, $h=10^{-6}$ (see Section~\ref{sec:sub:decrease}). This exclusion is several orders of magnitude smaller than competing methods.
\begin{itemize}
    \item FOSSIL~\cite{edwards2024fossil} synthesizes a network with square activation functions and verifies the resulting polynomial map within a box domain using the dReal SMT solver~\cite{gao2013dreal}. The $\delta$-satisfiability tolerance of dReal forces origin exclusion radius $\approx10^{-4}$.
    \item LyZNet~\cite{liu2024tool} couples quadratic and neural Lyapunov functions, employs Zubov's equation for ROA estimation, and uses dReal as its verifier (excludes radius $\approx10^{-2}$).
    \item Li et al.~\cite{li2025two} represent the state of the art, using a two-stage method to estimate the ROA via Zubov's equation and the GPU-accelerated $\alpha,\beta$-CROWN verifier. This approach requires an expensive pre-training, which we time separately. The final output of their Lyapunov candidate is scaled to $(0,1)$ via a sigmoid activation, and they exclude a re-scaled radius of $\approx 0.1$ around the origin.
\end{itemize}

\subsubsection{Results} Table~\ref{tab:CEGIS_table} reports the benchmarked systems alongside verification runtimes (in seconds).
Except for the smallest problems, Hyperlyve is consistently the fastest method, despite certifying larger domains with tighter numerical tolerances.
Hyperlyve demonstrates superior performance to SOTA when certifying sound and complete asymptotic stability for systems that have larger domains of interest (AKP-100), more complex dynamics (IPNN, Lorenz, and TLPRA), and more state dimensions (SIS 5+).

\subsection{Lyapunov Condition Verification}\label{sec:exp:verify}

As described in Section~\ref{sec:sub:decrease}, we use LiRPA~\cite{xu2020automatic} in an LP-tightened branch-and-bound scheme~\cite{yarbrough2026fitlrp} to compute Test~\ref{test:cond:dec}.
The obvious question is if the cost of Hyperlyve enumerating $V$ and then launching a verification procedure per region is superior to simply verifying the domain as a whole. Our results here emphatically demonstrate that enumeration is well-worth the cost. To explore this question, we benchmarked verification of $V$ against $\alpha,\! \beta$-CROWN~\cite{wang2021beta} and Vibecheck~\cite{bak2026vibecheck}, the first- and second-place competitors in 2026 VNN-COMP~\cite{brix2023vnncomp_overview}.

Specifically, we formulate Lyapunov Conditions~\ref{cond:pos_def} and~\ref{cond:dec}, the dynamics~\eqref{eq:original_system}, the candidate Lyapunov network $V$, and the domain constraints $\{\mathbf{x} \in \mathcal{B} \mid ||\mathbf{x}||_\infty \geq h \}$ as a single verification problem.
We set $\alpha,\! \beta$-CROWN to complete mode with projected gradient descent (PGD) enabled before bounding.
We report times for naive branching because we found them to outperform input-splitting. Complete clipping is enabled~\cite{zhou2025clip} (which substantially improved performance).
We report results for the enterprise-grade AMD MI210 GPU, but observed similar performance on an NVIDIA A100 GPU.

Table~\ref{tab:verify} reveals that Hyperlyve's parallelized enumeration and verification algorithm outperforms directly verifying $V$ without enumeration, even if direct verification leverages a powerful GPU.
The table shows verification results for both valid and invalid $V$ within the CEGIS loop.
 For some invalid networks $\alpha,\! \beta$-CROWN and Vibecheck each find a single easy counterexample and exit, while Hyperlyve continues to accumulate counterexamples from across the domain, which is more valuable for CEGIS.
 For example, on the early Lorenz run, all three tools find the first counterexample in approximately 2 seconds; however, Hyperlyve finds another 220 counterexamples from violating regions in the next 9 seconds. These points are added to the loss function~\eqref{eq:loss}, which helps $V$ learn at the hardest points of the domain.

\input{verify_table}

\section{Conclusion and Future Work}
\label{section:conclusion}
We developed a novel method for verifying deep neural Lyapunov functions with ReLU activations by leveraging geometric properties of hyperplane arrangements. We introduced and proved the validity of three tests that can be used to verify each Lyapunov condition separately. Formal proofs on the computational complexity of the neural Lyapunov function verification problem were also provided. We also introduced a region enumeration algorithm for deep ReLU neural networks. Furthermore, we formally introduced Hyperlyve and derived its time complexity. Through experiments, we illustrated the speedup Hyperlyve provides against other state-of-the-art neural Lyapunov function verifiers in terms of proving asymptotic stability of a closed-loop system.

Future work includes training optimal controllers while using Hyperlyve to guarantee asymptotic stability.
Additionally, bounding LiRPA and $\alpha,\! \beta$-CROWN inputs as polytopes instead of the current hyperrectanlge should be explored.

\bibliographystyle{IEEEtran}
\bibliography{mybib}

\end{document}

%% file: heaader.tex
\author{Brian Yarbrough$^*$,~\IEEEmembership{Student Member,~IEEE,} Jesse Wayment$^*$,~\IEEEmembership{Student Member,~IEEE,} Jingbo Wang, \\ Philip E. Par\'{e},~\IEEEmembership{Senior Member,~IEEE,} and Shreyas Sundaram,~\IEEEmembership{Senior Member,~IEEE}
\thanks{* Equal contribution. This work is supported in part by the US National Science Foundation (NSF-ECCS \#2238388, NSF CNS-2333487, NSF-CMMI \#2535574). Partial prior presentation in "Hyperlyve: Hyperplane Partitioning for Neural Lyapunov Verification" (CDC 2026, DOI forthcoming).}
\thanks{The authors are with the Elmore Family School of Electrical Engineering at Purdue University, West Lafayette, IN 47906, USA. E-mail: \{yarbroub, jwayment, wang6203, philpare, sundara2\}@purdue.edu.}
\thanks{Code and artifacts available at \url{https://github.com/Hyperlyve/hyperlyve.git}}}

%% file: cegis.tex
\begin{figure}[t]
    \centering
    \resizebox{\columnwidth}{!}{%
    \begin{tikzpicture}[
        >={Stealth[length=2.6mm]},
        thick,
        block/.style={rectangle, draw=black!55, rounded corners=3pt, fill=black!5,
            minimum width=3.0cm, minimum height=2.3cm, align=center, inner sep=4pt},
        verify/.style={block, draw=blue!70!black, line width=1.8pt, fill=blue!22, font=\bfseries},
        entry/.style={block, minimum width=3.0cm, minimum height=1.2cm, fill=white},
        decision/.style={diamond, draw=black!55, fill=white, aspect=1.6,
            align=center, inner sep=1pt, font=\small\bfseries},
        ok/.style={rectangle, draw=green!50!black, line width=1.4pt, rounded corners=3pt,
            fill=green!15, minimum width=2.6cm, minimum height=1.2cm, align=center,
            font=\bfseries, text=green!30!black},
        flow/.style={->, thick, black!75},
        lab/.style={font=\bfseries\small, fill=white, inner sep=1.5pt}
    ]

    \newcommand{\iconmesh}{\tikz[baseline=-0.5ex, scale=0.55]{
        \coordinate (a) at (-1.0,0.7); \coordinate (b) at (0.0,1.0); \coordinate (c) at (1.0,0.7);
        \coordinate (d) at (-0.4,0.0); \coordinate (e) at (0.6,0.1);
        \coordinate (f) at (-1.0,-0.7); \coordinate (g) at (0.0,-1.0); \coordinate (h) at (1.0,-0.7);
        \draw[thick, fill=blue!15] (a)--(b)--(d)--cycle;
        \draw[thick, fill=blue!30] (b)--(c)--(e)--(d)--cycle;
        \draw[thick, fill=blue!22] (c)--(h)--(e)--cycle;
        \draw[thick, fill=blue!38] (d)--(e)--(h)--(g)--cycle;
        \draw[thick, fill=blue!18] (a)--(d)--(g)--(f)--cycle;
    }}
    \newcommand{\iconcert}{\tikz[baseline=-0.5ex, scale=0.5]{
        \foreach \r/\f in {1.12/white, 0.74/blue!7, 0.38/blue!14}
            \draw[thick, blue!60!black, fill=\f] plot[domain=0:360, samples=90, smooth cycle] ({\r*(1+0.16*cos(3*\x))*cos(\x)}, {\r*(1+0.16*cos(3*\x))*sin(\x)});
        \draw[-{Stealth[length=1.8mm]}, red!70!black, line width=1pt] plot[domain=0:5.6, samples=40] ({0.98*exp(-0.36*\x)*cos(42*\x)}, {0.98*exp(-0.36*\x)*sin(42*\x)});
        \fill[red!70!black] (0.98,0) circle (0.08);
    }}
    \newcommand{\iconadd}{\tikz[baseline=-0.5ex, scale=0.55]{
        \draw[thick, fill=white] (-0.9,-0.6) rectangle (0.9,0.6);
        \draw[thick] (-0.9,0.2) -- (0.9,0.2) (-0.9,-0.2) -- (0.9,-0.2);
        \fill[red!70!black] (1.0,0.75) circle (0.42);
        \draw[white, line width=1.6pt] (0.75,0.75) -- (1.25,0.75) (1.0,0.5) -- (1.0,1.0);
    }}
    \newcommand{\iconsample}{\tikz[baseline=-0.5ex, scale=0.55]{
        \draw[thick, black!40, fill=white] (-1,-0.8) rectangle (1,0.8);
        \foreach \p in {(-0.6,0.3),(0.1,0.5),(0.6,0.1),(-0.2,-0.2),(0.4,-0.5),(-0.7,-0.5)}
            \fill[blue!70!black] \p circle (0.11);
        \foreach \p in {(0.7,0.55),(-0.05,-0.6),(-0.85,0.05)}
            \draw[orange!90!black, line width=1.2pt] \p circle (0.11);
    }}
    \newcommand{\icongd}{\tikz[baseline=-0.5ex, scale=0.55]{
        \draw[thick, domain=-1.15:1.15, samples=30] plot (\x, {0.9*\x*\x - 0.6});
        \fill[red!70!black] (0.95,0.0) circle (0.1);
        \fill[red!70!black] (0.55,-0.3) circle (0.1);
        \fill[red!70!black] (0.15,-0.57) circle (0.1);
        \draw[->, red!70!black, thick] (0.95,0.0) -- (0.6,-0.27);
        \draw[->, red!70!black, thick] (0.55,-0.3) -- (0.2,-0.54);
    }}

    \draw[black!50, dashed, line width=1pt, rounded corners=6pt, fill=black!4]
        (-1.8,2.15) -- (1.8,2.15) -- (1.8,-2.0) -- (6.0,-2.0) -- (6.0,-5.0) -- (-1.8,-5.0) -- cycle;
    \node[font=\small\bfseries\itshape, text=black!60, anchor=north]
        at (0.0,2.1) {Learner};

    \draw[blue!70!black, dashed, line width=1pt, rounded corners=6pt, fill=blue!4]
        (2.4,-1.4) rectangle (10.2,2.15);
    \node[font=\small\bfseries\itshape, text=blue!60!black, anchor=north]
        at (6.3,2.1) {Hyperlyve Verifier};

    \node[block]  (train) at (0,0)   {\icongd\\[4pt] Train\\ReLU network};
    \node[verify] (poly)  at (4.2,0) {\iconmesh\\[4pt] Polyhedral\\Enumeration};
    \node[verify] (cert)  at (8.4,0) {\iconcert\\[4pt] Lyapunov\\Certification};
    \node[block]    (sample) at (0,-3.6)   {\iconsample\\[4pt] Sample region\\of interest};
    \node[block]    (addce)  at (4.2,-3.6) {\iconadd\\[4pt] Add counter-\\examples};
    \node[decision] (check)  at (8.4,-3.6) {Counter-\\example?};
    \node[entry] (target) at (0,-6.3) {Dynamics $f$\\Domain $\mathcal{B}$};
    \node[ok] (verified) at (8.4,-6.3) {\tikz[baseline=-0.5ex]{\draw[green!40!black, line width=2.4pt, line cap=round, line join=round] (0,0.02) -- (0.16,-0.17) -- (0.5,0.3);}~Verified};
    \node[font=\bfseries, text=black!80, align=center] at (4.2,-6.3)
        {Proving Asymptotic Stability\\with Hyperlyve};

    \draw[flow] (target) -- (sample);
    \draw[flow] (sample) -- (train);
    \draw[flow] (train) -- (poly);
    \draw[flow] (poly) -- (cert);
    \draw[flow] (cert) -- (check);
    \draw[flow] (check.south) -- node[lab, pos=0.4] {No} (verified.north);
    \draw[flow] (check.west) -- node[lab, pos=0.4] {Yes} (addce.east);
    \draw[flow] (addce) -- (sample);

    \end{tikzpicture}%
    }
    \caption{The Hyperlyve counterexample-guided inductive synthesis (CEGIS) framework. A candidate ReLU neural Lyapunov function $V$ is trained by a Learner (gray) on sampled states of $f$ in the domain $\mathcal{B}$ before being passed to Hyperlyve verifier (blue). Hyperlyve enumerates the induced activation regions of $V$ and checks Lyapunov stability conditions across each region in parallel. Hyperlyve returns counterexamples to the Learner, which adds them to the dataset and completes another round of training. This CEGIS loop continues until no counterexamples are found for $V$, certifying the asymptotic stability of the nonlinear dynamics over the region of interest.}
    \label{fig:cegis_loop}
\end{figure}

%% file: lyapunov_table.tex
\begin{table*}[t]
\renewcommand{\arraystretch}{1.3}
\caption{Comparison of nonlinear asymptotic stability certification frameworks. Hyperlyve outperforms state-of-the-art methods on systems that have larger domains of interest (AKP-100), more complex dynamics (IPNN, Lorenz, and TLPRA), and more state dimensions (SIS 5+).}
\label{tab:CEGIS_table}
\centering
\setlength{\tabcolsep}{8pt}
\resizebox{\textwidth}{!}{%
\begin{threeparttable}
\begin{tabular}{lr cc cc ccc ccc}
\toprule
& & \multicolumn{2}{c}{Fossil~\cite{edwards2024fossil}} & \multicolumn{2}{c}{LyZNet~\cite{liu2024tool}} & \multicolumn{3}{c}{Li et al.~\cite{li2025two}} & \multicolumn{3}{c}{\textbf{Hyperlyve}} \\
\cmidrule(lr){3-4}\cmidrule(lr){5-6}\cmidrule(lr){7-9}\cmidrule(lr){10-12}
System & $p$ & Neurons & $T$ (s) & Neurons & $T$ (s) & Neurons & $T_\mathrm{pre}$ (s) & $T_\mathrm{cert}$ (s) & Neurons & Regions & $T$ (s) \\
\midrule
AKP-10 & 2 & [6] & 11.6 & [30,30] & 225 & [40,40] & 258 & 152 & [8,8] & 73 & \textbf{8.3} \\
AKP-100 & 2 & [6] & \TO & [30,30] & 175 & [40,40] & \TO & -- & [16,16] & 246 & \textbf{19} \\
IPNN & 2 & [5] & 703 & [30,30] & 318 & [40,40] & 395 & 200 & [8,8] & 37 & \textbf{11} \\
Lorenz & 3 & [10] & \TO & [16,16] & 251 & [64,64] & 269 & 156 & [24,24] & 10,868 & \textbf{43} \\
TLPRA & 4 & [10] & \TO & -- & \TO & [64,64] & 366 & \ERR & [14,14,14] & 3,010 & \textbf{3194} \\
\midrule
SIS & 3 & [10] & \textbf{2.62} & [10,10] & 150 & -- & -- & -- & [8,8,8] & 85 & 7.0 \\
SIS & 4 & [10] & \textbf{3.07} & [10,10] & \TO & -- & -- & -- & [8,8,8] & 167 & 6.3 \\
SIS & 5 & [10] & 112 & -- & -- & -- & -- & -- & [8,8,8] & 46 & \textbf{10.3} \\
SIS & 6 & [10] & 111$^{4/5}$ & -- & -- & -- & -- & -- & [8,8,8] & 256 & \textbf{8.3} \\
SIS & 7 & [10] & 642$^{4/5}$ & -- & -- & -- & -- & -- & [8,8,8] & 131 & \textbf{13} \\
SIS & 8 & [10] & \TO & -- & -- & -- & -- & -- & [8,8,8] & 164 & \textbf{30} \\
SIS & 9 & [10] & \TO & -- & -- & -- & -- & -- & [8,8,8] & 184 & \textbf{17} \\
SIS & 10 & [10] & \TO & -- & -- & -- & -- & -- & [8,8,8] & 107 & \textbf{77} \\
\bottomrule
\end{tabular}
\begin{tablenotes}\footnotesize
    \item[] ``Neurons'' denotes the per-layer number of neurons. Network sizes are set to match the closest analogous systems in the respective original papers; differences in activation functions and training procedures discourage exact structural matching. FOSSIL and Hyperlyve certify the entire given domain $\mathcal{B}$, whereas LyZNet and Li et al. certify a discovered region of attraction (ROA). The first stage of Li et al.'s method estimates the ROA, so we separate pre-computation ($T_\mathrm{pre}$) from verification time ($T_\mathrm{cert}$). Times are averages of converged runs. Bold values indicate the fastest time per row.
    \item[\TO] Denotes a 1-hour timeout limit.
    \item[--] Unattempted run due to previous timeouts (Fossil, LyZNet) or inability to handle offset domains (Li et al. with SIS).
    \item[\ERR] Li et al. error on TLPRA stems from certifying an infinitesimal domain on two runs and producing no certificate on others.
    \item[$^{4/5}$] Fraction of runs that converged to valid certificates.
\end{tablenotes}
\end{threeparttable}%
}
\end{table*}

%% file: verify_table.tex
\begin{table}[!t]
\centering
\footnotesize
\caption{Verification and counterexample generation for CPA neural Lyapunov functions. Hyperlyve's per-region verification outperforms GPU-based verifiers on valid functions, completing checks in seconds where $\alpha,\!\beta$-CROWN~\cite{wang2021beta} and Vibecheck~\cite{bak2026vibecheck} hit the 1-hour timeout. While Hyperlyve finds the first counterexample in comparable time to competing tools, it uniquely identifies a counterexample in every violating activation region, which improves subsequent CEGIS training.}\label{tab:verify}
\begin{threeparttable}

\begin{subtable}{\linewidth}
\centering
\caption{Valid Lyapunov functions}
\label{tab:verify_valid}
\begin{tabular*}{\linewidth}{@{\extracolsep{\fill}}l r r r r r@{}}
\toprule
\rlap{System}\hphantom{\begin{tabular}{@{}l@{}}AKP-10\\AKP-100\\IPNN\\Lorenz\\TLPRA\\SIS-3\\SIS-4\\SIS-5\\SIS-6\\SIS-7$\ldots$10\\AKP-10$^\mathrm{e}$\\AKP-10$^\mathrm{l}$\\AKP-100$^\mathrm{e}$\\AKP-100$^\mathrm{l}$\\IPNN$^\mathrm{e}$\\IPNN$^\mathrm{l}$\\Lorenz$^\mathrm{e}$\\Lorenz$^\mathrm{l}$\\TLPRA$^\mathrm{e}$\\TLPRA$^\mathrm{l}$\\System\end{tabular}} & \hphantom{\begin{tabular}{@{}r@{}}9.8\\10.9\\14.7\\32.7\\\TO\\13.4\\15.5\\139.7\\21.8\\\TO\\\textbf{1.8}\\\textbf{2.0}\\2.3\\7.1\\\textbf{1.8}\\2.1\\\textbf{2.0}\\47.3\\2.6\\3159.7\\$\alpha,\!\beta$-CROWN\end{tabular}}\llap{$\alpha,\!\beta$-CROWN} & \hphantom{\begin{tabular}{@{}r@{}}5.6\\6.4\\6.9\\\textbf{8.4}\\\TO\\7.6\\\TO\\--\\--\\--\\2.0\\2.2\\\textbf{2.1}\\\textbf{2.1}\\2.3\\2.2\\2.3\\8.8\\\textbf{2.3}\\\TO\\Vibecheck\end{tabular}}\llap{Vibecheck} & \hphantom{\begin{tabular}{@{}r@{}}\textbf{3.1}\\\textbf{3.3}\\\textbf{6.5}\\11.7\\\textbf{143.2}\\\textbf{3.3}\\\textbf{3.2}\\\textbf{3.2}\\\textbf{3.4}\\\textbf{3.2--3.3}\\1.9\\\textbf{2.0}\\2.3\\2.6\\2.0\\\textbf{1.9}\\2.4\\\textbf{2.7}\\2.4\\\textbf{13.2}\\Hyperlyve\end{tabular}}\llap{Hyperlyve} & \multicolumn{2}{c@{}}{\hphantom{Hyperlyve total}} \\
\midrule
AKP-10 & 9.8 & 5.6 & \textbf{3.1} & \hphantom{\begin{tabular}{@{}r@{}}3.0\\3.0\\3.6\\3.6\\5.0\\5.7\\11.8\\11.8\\136.2\\124.9\\Time\end{tabular}} & \hphantom{\begin{tabular}{@{}r@{}}5\\2\\7\\1\\7\\2\\221\\1\\545\\1\\\#CEx\end{tabular}} \\
AKP-100 & 10.9 & 6.4 & \textbf{3.3} &  &  \\
IPNN & 14.7 & 6.9 & \textbf{6.5} &  &  \\
Lorenz & 32.7 & \textbf{8.4} & 11.7 &  &  \\
TLPRA & \TO & \TO & \textbf{143.2} &  &  \\
SIS-3 & 13.4 & 7.6 & \textbf{3.3} &  &  \\
SIS-4 & 15.5 & \TO & \textbf{3.2} &  &  \\
SIS-5 & 139.7 & \TO & \textbf{3.2} &  &  \\
SIS-6 & 21.8 & \TO & \textbf{3.4} &  &  \\
SIS-7$\ldots$10 & \TO & \TO & \textbf{3.2--3.3} &  &  \\
\bottomrule
\end{tabular*}
\end{subtable}

\medskip
\begin{subtable}{\linewidth}
\centering
\caption{Invalid Lyapunov functions (rejected during CEGIS)}
\label{tab:verify_invalid}
\begin{tabular*}{\linewidth}{@{\extracolsep{\fill}}l r r r r r@{}}
\toprule
 & \multicolumn{3}{c}{First counterexample} & \multicolumn{2}{c@{}}{Hyperlyve total} \\
\cmidrule(lr){2-4}\cmidrule(l){5-6}
\rlap{System}\hphantom{\begin{tabular}{@{}l@{}}AKP-10\\AKP-100\\IPNN\\Lorenz\\TLPRA\\SIS-3\\SIS-4\\SIS-5\\SIS-6\\SIS-7$\ldots$10\\AKP-10$^\mathrm{e}$\\AKP-10$^\mathrm{l}$\\AKP-100$^\mathrm{e}$\\AKP-100$^\mathrm{l}$\\IPNN$^\mathrm{e}$\\IPNN$^\mathrm{l}$\\Lorenz$^\mathrm{e}$\\Lorenz$^\mathrm{l}$\\TLPRA$^\mathrm{e}$\\TLPRA$^\mathrm{l}$\\System\end{tabular}} & \hphantom{\begin{tabular}{@{}r@{}}9.8\\10.9\\14.7\\32.7\\\TO\\13.4\\15.5\\139.7\\21.8\\\TO\\\textbf{1.8}\\\textbf{2.0}\\2.3\\7.1\\\textbf{1.8}\\2.1\\\textbf{2.0}\\47.3\\2.6\\3159.7\\$\alpha,\!\beta$-CROWN\end{tabular}}\llap{$\alpha,\!\beta$-CROWN} & \hphantom{\begin{tabular}{@{}r@{}}5.6\\6.4\\6.9\\\textbf{8.4}\\\TO\\7.6\\\TO\\--\\--\\--\\2.0\\2.2\\\textbf{2.1}\\\textbf{2.1}\\2.3\\2.2\\2.3\\8.8\\\textbf{2.3}\\\TO\\Vibecheck\end{tabular}}\llap{Vibecheck} & \hphantom{\begin{tabular}{@{}r@{}}\textbf{3.1}\\\textbf{3.3}\\\textbf{6.5}\\11.7\\\textbf{143.2}\\\textbf{3.3}\\\textbf{3.2}\\\textbf{3.2}\\\textbf{3.4}\\\textbf{3.2--3.3}\\1.9\\\textbf{2.0}\\2.3\\2.6\\2.0\\\textbf{1.9}\\2.4\\\textbf{2.7}\\2.4\\\textbf{13.2}\\Hyperlyve\end{tabular}}\llap{Hyperlyve} & \hphantom{\begin{tabular}{@{}r@{}}3.0\\3.0\\3.6\\3.6\\5.0\\5.7\\11.8\\11.8\\136.2\\124.9\\Time\end{tabular}}\llap{Time} & \hphantom{\begin{tabular}{@{}r@{}}5\\2\\7\\1\\7\\2\\221\\1\\545\\1\\\#CEx\end{tabular}}\llap{\#CEx} \\
\midrule
AKP-10$^\mathrm{e}$ & \textbf{1.8} & 2.0 & 1.9 & 3.0 & 5 \\
AKP-10$^\mathrm{l}$ & \textbf{2.0} & 2.2 & \textbf{2.0} & 3.0 & 2 \\
AKP-100$^\mathrm{e}$ & 2.3 & \textbf{2.1} & 2.3 & 3.6 & 7 \\
AKP-100$^\mathrm{l}$ & 7.1 & \textbf{2.1} & 2.6 & 3.6 & 1 \\
IPNN$^\mathrm{e}$ & \textbf{1.8} & 2.3 & 2.0 & 5.0 & 7 \\
IPNN$^\mathrm{l}$ & 2.1 & 2.2 & \textbf{1.9} & 5.7 & 2 \\
Lorenz$^\mathrm{e}$ & \textbf{2.0} & 2.3 & 2.4 & 11.8 & 221 \\
Lorenz$^\mathrm{l}$ & 47.3 & 8.8 & \textbf{2.7} & 11.8 & 1 \\
TLPRA$^\mathrm{e}$ & 2.6 & \textbf{2.3} & 2.4 & 136.2 & 545 \\
TLPRA$^\mathrm{l}$ & 3159.7 & \TO & \textbf{13.2} & 124.9 & 1 \\
\bottomrule
\end{tabular*}
\end{subtable}

\begin{tablenotes}\footnotesize
    \item[] All times in seconds. Systems, domains, and architectures match Table~\ref{tab:CEGIS_table}, and exclude $||x||_\infty < h$ around the origin.
    \item[] SIS-7$\ldots$10 represents the range across instances. 
    \item[] \textit{First counterexample} measures time until the check of the first region with a counterexample returns. \textit{Hyperlyve total} is total time to check all regions, and \textit{\#CEx} tracks the number of counterexamples.
    \item[$\mathrm{e}$/$\mathrm{l}$] early/late CEGIS iterations, where late is closer to a valid function.
    \item[\TO] Denotes a 60-minute timeout limit.
\end{tablenotes}

\end{threeparttable}
\end{table}

%% file: mybib.bib
@inproceedings{anon2026hyperlyve,
  author    = {Anonymous},
  title     = {{Hyperlyve}: Hyperplane Partitioning for Neural Lyapunov Verification },
  booktitle = {Proc. 65th IEEE Conference on Decision and Control (CDC)},
  address   = {Honolulu, HI, USA},
  month     = dec,
  year      = {2026},
  note      = {to appear; arXiv:2605.03992, doi:10.48550/arXiv.2605.03992}
}

@article{yarbrough2026fitlrp,
  title   = {{Fit-LRP}: Fitting Linear Relaxations to Polytopes for Neural Network Verification over Non-Convex Regions},
  author  = {Yarbrough, Brian and Wayment, Jesse and Wang, Jingbo and Par{\'e}, Philip E. and Sundaram, Shreyas},
  journal = {arXiv preprint arXiv:2611.16003},
  year    = {2026}
}

@article{robey2026beyond,
  author  = {Robey, Alexander and Ravichandran, Zachary and Jones, Eliot Krzysztof and Perlo, Jared and Barez, Fazl and Kumar, Vijay and Kolter, J. Zico and Hassani, Hamed and Pappas, George J.},
  title   = {Beyond alignment: Why robotic foundation models need context-aware safety},
  journal = {Sci. Robot.},
  volume  = {11},
  number  = {113},
  pages   = {eaef2191},
  year    = {2026},
  doi     = {10.1126/scirobotics.aef2191}
}

@inproceedings{ahmadi2012difficulty,
  author    = {Ahmadi, Amir Ali},
  title     = {On the difficulty of deciding asymptotic stability of cubic homogeneous vector fields},
  booktitle = {Proc. Amer. Control Conf. (ACC)},
  pages     = {3334--3339},
  year      = {2012}
}

@article{SamanipourTAC,
  author={Samanipour, Pouya and Poonawala, Hasan A.},
  journal={IEEE Transactions on Automatic Control}, 
  title={Stability Analysis and Controller Synthesis Using Single-Hidden-Layer ReLU Neural Networks}, 
  year={2024},
  volume={69},
  number={1},
  pages={202-213},
  doi={10.1109/TAC.2023.3270219}
}

@inproceedings{SamanipouraAllerton,
  author={Samanipour, Pouya and Poonawala, Hasan A.},
  booktitle={2023 59th Annual Allerton Conference on Communication, Control, and Computing (Allerton)}, 
  title={Automated Stability Analysis of Piecewise Affine Dynamics Using Vertices}, 
  year={2023},
  volume={},
  number={},
  pages={1-8},
  doi={10.1109/Allerton58177.2023.10313502}
}

@phdthesis{samanipouranalysis,
  author = {Samanipour, Pouya},
  title  = {Analysis of closed-loop dynamical systems using {ReLU} {NNs}},
  school = {Univ. Kentucky},
  address = {Lexington, KY, USA},
  type   = {Ph.D. dissertation},
  year   = {2025},
  doi    = {10.13023/etd.2025.461}
}

@inproceedings{liu2024tool,
  author    = {Liu, Jun and Meng, Yiming and Fitzsimmons, Maxwell and Zhou, Ruikun},
  title     = {Tool {LyzNet}: A lightweight {Python} tool for learning and verifying neural {Lyapunov} functions and regions of attraction},
  booktitle = {Proc. 27th ACM Int. Conf. Hybrid Syst., Comput. Control (HSCC)},
  pages     = {1--8},
  year      = {2024}
}

@inproceedings{liu_learning_max_lyapunov,
  author    = {Liu, Jun and Meng, Yiming and Fitzsimmons, Maxwell and Zhou, Ruikun},
  title     = {Towards learning and verifying maximal neural {Lyapunov} functions},
  booktitle = {Proc. 62nd IEEE Conference on Decision and Control (CDC)},
  pages     = {8012--8019},
  year      = {2023},
  doi       = {10.1109/CDC49753.2023.10383299}
}

@inproceedings{huang_MNLF,
  author    = {Huang, Junyue and Li, Shaoyuan and Yin, Xiang},
  title     = {Stability verification for switched systems using neural multiple {Lyapunov} functions},
  booktitle = {Proc. 64th IEEE Conference on Decision and Control (CDC)},
  pages     = {299--304},
  year      = {2025},
  doi       = {10.1109/CDC57313.2025.11312529}
}

@inproceedings{liu2024compositionally,
  author    = {Liu, Jun and Meng, Yiming and Fitzsimmons, Maxwell and Zhou, Ruikun},
  title     = {Compositionally verifiable vector neural {Lyapunov} functions for stability analysis of interconnected nonlinear systems},
  booktitle = {Proc. Amer. Control Conf. (ACC)},
  pages     = {4789--4794},
  year      = {2024}
}

@inproceedings{chen2021learning,
  author    = {Chen, Shaoru and Fazlyab, Mahyar and Morari, Manfred and Pappas, George J. and Preciado, Victor M.},
  title     = {Learning region of attraction for nonlinear systems},
  booktitle = {Proc. 60th IEEE Conference on Decision and Control (CDC)},
  pages     = {6477--6484},
  year      = {2021}
}

@inproceedings{dai2020counter,
  author    = {Dai, Hongkai and Landry, Benoit and Pavone, Marco and Tedrake, Russ},
  title     = {Counter-example guided synthesis of neural network {Lyapunov} functions for piecewise linear systems},
  booktitle = {Proc. 59th IEEE Conference on Decision and Control (CDC)},
  pages     = {1274--1281},
  year      = {2020}
}

@inproceedings{yang2024lyapunov,
  author    = {Yang, Lujie and Dai, Hongkai and Shi, Zhouxing and Hsieh, Cho-Jui and Tedrake, Russ and Zhang, Huan},
  title     = {{Lyapunov}-stable neural control for state and output feedback: A novel formulation},
  booktitle = {Proc. 41st Int. Conf. Mach. Learn. (ICML)},
  pages     = {56033--56046},
  year      = {2024}
}

@inproceedings{li2025two,
    author = {Li, Haoyu and Zhong, Xiangru and Hu, Bin and Zhang, Huan},
     booktitle = {Advances in Neural Information Processing Systems},
     doi = {10.52202/085713-2480},
     editor = {D. Belgrave and C. Zhang and H. Lin and R. Pascanu and P. Koniusz and M. Ghassemi and N. Chen},
     pages = {73980--74009},
     publisher = {Curran Associates, Inc.},
     title = {Two‑Stage Learning of Stabilizing Neural Controllers via {Z}ubov Sampling and Iterative Domain Expansion},
     volume = {38, Main Conference},
     year = {2025}
}

@article{moulay2010some,
  author  = {Moulay, Emmanuel},
  title   = {Some properties of {Lyapunov} function sets},
  journal = {Proc. Amer. Math. Soc.},
  volume  = {138},
  number  = {11},
  pages   = {4067--4073},
  year    = {2010}
}

@book{dantzig2016linear,
  author    = {Dantzig, George B.},
  title     = {{Linear Programming and Extensions}},
  publisher = {Princeton Univ. Press},
  address   = {Princeton, NJ, USA},
  year      = {2016}
}

@book{khalil2002nonlinear,
  author    = {Khalil, Hassan K.},
  title     = {{Nonlinear Systems}},
  edition   = {3rd},
  publisher = {Prentice-Hall},
  address   = {Upper Saddle River, NJ, USA},
  year      = {2002}
}

@inproceedings{gaby2022Lyapunov_net,
  author    = {Gaby, Nathan and Zhang, Fumin and Ye, Xiaojing},
  title     = {{Lyapunov}-net: A deep neural network architecture for {Lyapunov} function approximation},
  booktitle = {Proc. 61st IEEE Conference on Decision and Control (CDC)},
  pages     = {2091--2096},
  year      = {2022}
}

@incollection{stanley2007introduction,
  author    = {Stanley, Richard P.},
  title     = {An introduction to hyperplane arrangements},
  booktitle = {Geometric Combinatorics},
  volume    = {13},
  pages     = {389--496},
  publisher = {American Mathematical Society},
  year      = {2007},
  doi       = {10.1090/pcms/013/08}
}

@inproceedings{ahmadi2011globally,
  author    = {Ahmadi, Amir Ali and Krstic, Miroslav and Parrilo, Pablo A.},
  title     = {A globally asymptotically stable polynomial vector field with no polynomial {Lyapunov} function},
  booktitle = {Proc. 50th IEEE Conf. Decis. Control Eur. Control Conf. (CDC-ECC)},
  pages     = {7579--7580},
  year      = {2011},
  doi       = {10.1109/CDC.2011.6161499}
}

@article{lajmanovich1976deterministic,
  author  = {Lajmanovich, Ana and Yorke, James A.},
  title   = {A deterministic model for gonorrhea in a nonhomogeneous population},
  journal = {Math. Biosci.},
  volume  = {28},
  number  = {3--4},
  pages   = {221--236},
  year    = {1976}
}

@article{erdos1959random,
  author  = {Erd{\H{o}}s, Paul and R{\'e}nyi, Alfr{\'e}d},
  title   = {On random graphs {I}},
  journal = {Publ. Math. Debrecen},
  volume  = {6},
  pages   = {290--297},
  year    = {1959}
}

@techreport{kraft1988software,
    author      = {Kraft, Dieter},
    title       = {A software package for sequential quadratic programming},
    institution = {Inst. Dyn. Flugsysteme, Deutsche Forschungs- und Versuchsanstalt f{\"u}r Luft- und Raumfahrt (DFVLR)},
    address     = {Oberpfaffenhofen, Germany},
    number      = {DFVLR-FB 88-28},
    year        = {1988}
}

@article{murty1987some,
  author  = {Murty, Katta G. and Kabadi, Santosh N.},
  title   = {Some {NP}-complete problems in quadratic and nonlinear programming},
  journal = {Math. Program.},
  volume  = {39},
  number  = {2},
  pages   = {117--129},
  year    = {1987}
}

@inproceedings{wang2021beta,
   author = {Wang, Shiqi and Zhang, Huan and Xu, Kaidi and Lin, Xue and Jana, Suman and Hsieh, Cho-Jui and Kolter, J. Zico},
 booktitle = {Advances in Neural Information Processing Systems},
 editor = {M. Ranzato and A. Beygelzimer and Y. Dauphin and P.S. Liang and J. Wortman Vaughan},
 pages = {29909--29921},
 publisher = {Curran Associates, Inc.},
 title = {Beta-CROWN: Efficient Bound Propagation with Per-neuron Split Constraints for Neural Network Robustness Verification},
 volume = {34},
 year = {2021}
}

@inproceedings{zhou2022neural,
author = {Zhou, Ruikun and Quartz, Thanin and De Sterck, Hans and Liu, Jun},
 booktitle = {Advances in Neural Information Processing Systems},
 doi = {10.52202/068431-2111},
 editor = {S. Koyejo and S. Mohamed and A. Agarwal and D. Belgrave and K. Cho and A. Oh},
 pages = {29113--29125},
 publisher = {Curran Associates, Inc.},
 title = {Neural Lyapunov Control of Unknown Nonlinear Systems with Stability Guarantees},
 volume = {35},
 year = {2022}
}

@inproceedings{richards2018lyapunov,
  author    = {Richards, Spencer M. and Berkenkamp, Felix and Krause, Andreas},
  title     = {The {Lyapunov} neural network: Adaptive stability certification for safe learning of dynamical systems},
  booktitle = {Proc. Conf. Robot Learn. (CoRL)},
  pages     = {466--476},
  year      = {2018}
}

@inproceedings{wu2023neural,
   author = {Wu, Junlin and Clark, Andrew and Kantaros, Yiannis and Vorobeychik, Yevgeniy},
 booktitle = {Advances in Neural Information Processing Systems},
 doi = {10.52202/075280-0131},
 editor = {A. Oh and T. Naumann and A. Globerson and K. Saenko and M. Hardt and S. Levine},
 pages = {2939--2955},
 publisher = {Curran Associates, Inc.},
 title = {Neural Lyapunov Control for Discrete-Time Systems},
 volume = {36},
 year = {2023}
}

@inproceedings{petridis2006construction,
  author    = {Petridis, Vassilios and Petridis, Stavros},
  title     = {Construction of neural network based {Lyapunov} functions},
  booktitle = {Proc. IEEE Int. Joint Conf. Neural Netw. (IJCNN)},
  pages     = {5059--5065},
  year      = {2006}
}

@article{barber1996quickhull,
  author  = {Barber, C. Bradford and Dobkin, David P. and Huhdanpaa, Hannu},
  title   = {The quickhull algorithm for convex hulls},
  journal = {ACM Trans. Math. Softw.},
  volume  = {22},
  number  = {4},
  pages   = {469--483},
  year    = {1996}
}

@article{valiant1979complexity,
  author  = {Valiant, Leslie G.},
  title   = {The complexity of computing the permanent},
  journal = {Theor. Comput. Sci.},
  volume  = {8},
  number  = {2},
  pages   = {189--201},
  year    = {1979}
}

@inproceedings{gao2013dreal,
  author    = {Gao, Sicun and Kong, Soonho and Clarke, Edmund M.},
  title     = {{dReal}: An {SMT} solver for nonlinear theories over the reals},
  booktitle = {Proc. Int. Conf. Autom. Deduction (CADE)},
  pages     = {208--214},
  year      = {2013}
}

@inproceedings{stargalla2025computational,
   author = {Stargalla, Moritz and Hertrich, Christoph and Reichman, Daniel},
     booktitle = {Advances in Neural Information Processing Systems},
     doi = {10.52202/085713-4199},
     editor = {D. Belgrave and C. Zhang and H. Lin and R. Pascanu and P. Koniusz and M. Ghassemi and N. Chen},
     pages = {125970--125988},
     publisher = {Curran Associates, Inc.},
     title = {The Computational Complexity of Counting Linear Regions in ReLU Neural Networks},
     volume = {38, Main Conference},
     year = {2025}
    }

@inproceedings{katz2017reluplex,
  author    = {Katz, Guy and Barrett, Clark and Dill, David L. and Julian, Kyle and Kochenderfer, Mykel J.},
  title     = {Reluplex: An efficient {SMT} solver for verifying deep neural networks},
  booktitle = {Proc. Int. Conf. Comput. Aided Verification (CAV)},
  pages     = {97--117},
  year      = {2017}
}

@article{TNNLSXiangReachbility2,
  author  = {Xiang, Weiming and Tran, Hoang-Dung and Yang, Xiaodong and Johnson, Taylor T.},
  title   = {Reachable set estimation for neural network control systems: A simulation-guided approach},
  journal = {IEEE Trans. Neural Netw. Learn. Syst.},
  volume  = {32},
  number  = {5},
  pages   = {1821--1830},
  year    = {2021},
  doi     = {10.1109/TNNLS.2020.2991090}
}

@article{TNNLSXiangReachability1,
  author  = {Xiang, Weiming and Tran, Hoang-Dung and Johnson, Taylor T.},
  title   = {Output reachable set estimation and verification for multilayer neural networks},
  journal = {IEEE Trans. Neural Netw. Learn. Syst.},
  volume  = {29},
  number  = {11},
  pages   = {5777--5783},
  year    = {2018},
  doi     = {10.1109/TNNLS.2018.2808470}
}

@article{vincent2025tnnls,
  author={Vincent, Joseph A. and Schwager, Mac},
  journal={IEEE Transactions on Neural Networks and Learning Systems}, 
  title={Reachable Polyhedral Marching (RPM): An Exact Analysis Tool for Deep-Learned Control Systems}, 
  year={2025},
  volume={36},
  number={10},
  pages={19225-19239},
  doi={10.1109/TNNLS.2025.3571720}
}

@inproceedings{villani2024pice,
  author    = {Villani, Mattia Jacopo and Albini, Emanuele and Sharma, Shubham and Mishra, Saumitra and Amoukou, Salim Ibrahim and Magazzeni, Daniele and Veloso, Manuela},
  title     = {{PICE}: Polyhedral complex informed counterfactual explanations},
  booktitle = {Proc. AAAI/ACM Conf. AI, Ethics, Soc.},
  volume    = {7},
  number    = {1},
  pages     = {1502--1513},
  year      = {2024}
}

@inproceedings{berzins2023edge,
  author    = {Berzins, Arturs},
  title     = {Polyhedral complex extraction from {ReLU} networks using edge subdivision},
  booktitle = {Proc. 40th Int. Conf. Mach. Learn. (ICML)},
  series    = {Proc. Mach. Learn. Res.},
  volume    = {202},
  year      = {2023}
}

@inproceedings{humayun2022splinecam,
  author    = {Humayun, Ahmed Imtiaz and Balestriero, Randall and Balakrishnan, Guha and Baraniuk, Richard},
  title     = {{SplineCam}: Exact visualization and characterization of deep network geometry and decision boundaries},
  booktitle = {Proc. IEEE/CVF Conf. Comput. Vis. Pattern Recognit. (CVPR)},
  year      = {2023}
}

@article{masden2025algorithmic,
  author  = {Masden, Marissa},
  title   = {Algorithmic determination of the combinatorial structure of the linear regions of {ReLU} neural networks},
  journal = {SIAM J. Appl. Algebra Geom.},
  volume  = {9},
  number  = {2},
  pages   = {374--404},
  year    = {2025}
}

@inproceedings{xu2021traversing,
  author    = {Xu, Shaojie and Vaughan, Joel and Chen, Jie and Zhang, Aijun and Sudjianto, Agus},
  title     = {Traversing the local polytopes of {ReLU} neural networks},
  booktitle = {Proc. AAAI-22 Workshop Adversarial Mach. Learn. Beyond},
  year      = {2021}
}

@inproceedings{gaines2026characterizing,
  title     = {Characterizing the Discrete Geometry of {ReLU} Networks},
  author    = {Blake B. Gaines and Jinbo Bi},
  booktitle = {International Conference on Learning Representations (ICLR)},
  year      = {2026},
}

@inproceedings{xu2020automatic,
 author = {Xu, Kaidi and Shi, Zhouxing and Zhang, Huan and Wang, Yihan and Chang, Kai-Wei and Huang, Minlie and Kailkhura, Bhavya and Lin, Xue and Hsieh, Cho-Jui},
 booktitle = {Advances in Neural Information Processing Systems},
 editor = {H. Larochelle and M. Ranzato and R. Hadsell and M.F. Balcan and H. Lin},
 pages = {1129--1141},
 publisher = {Curran Associates, Inc.},
 title = {Automatic Perturbation Analysis for Scalable Certified Robustness and Beyond},
 volume = {33},
 year = {2020}
}

@article{salzer2023reachability,
  author  = {S{\"a}lzer, Marco and Lange, Martin},
  title   = {Reachability in simple neural networks},
  journal = {Fundamenta Informaticae},
  volume  = {189},
  number  = {3--4},
  pages   = {241--259},
  year    = {2023},
  doi     = {10.3233/FI-222160}
}

@inproceedings{edwards2024fossil,
  author    = {Edwards, Alec and Peruffo, Andrea and Abate, Alessandro},
  title     = {{Fossil} 2.0: Formal certificate synthesis for the verification and control of dynamical models},
  booktitle = {Proc. 27th ACM Int. Conf. Hybrid Syst., Comput. Control (HSCC)},
  year      = {2024}
}

@article{dai2021lyapunov,
  author  = {Dai, Hongkai and Landry, Benoit and Yang, Lujie and Pavone, Marco and Tedrake, Russ},
  title   = {{Lyapunov}-stable neural-network control},
  journal = {arXiv preprint arXiv:2109.14152},
  year    = {2021}
}

@inproceedings{hanin2019deep,
 author = {Hanin, Boris and Rolnick, David},
 booktitle = {Advances in Neural Information Processing Systems},
 editor = {H. Wallach and H. Larochelle and A. Beygelzimer and F. d\textquotesingle Alch\'{e}-Buc and E. Fox and R. Garnett},
 pages = {},
 publisher = {Curran Associates, Inc.},
 title = {Deep ReLU Networks Have Surprisingly Few Activation Patterns},
 volume = {32},
 year = {2019}
}

@inproceedings{chang2019neural,
 author = {Chang, Ya-Chien and Roohi, Nima and Gao, Sicun},
 booktitle = {Advances in Neural Information Processing Systems},
 editor = {H. Wallach and H. Larochelle and A. Beygelzimer and F. d\textquotesingle Alch\'{e}-Buc and E. Fox and R. Garnett},
 pages = {},
 publisher = {Curran Associates, Inc.},
 title = {Neural {L}yapunov Control},
 volume = {32},
 year = {2019}
}

@article{mcmullen1970maximum,
  author  = {McMullen, Peter},
  title   = {The maximum numbers of faces of a convex polytope},
  journal = {Mathematika},
  volume  = {17},
  number  = {2},
  pages   = {179--184},
  year    = {1970}
}

@article{lorenz1963deterministic,
  author  = {Lorenz, Edward N.},
  title   = {Deterministic nonperiodic flow},
  journal = {J. Atmos. Sci.},
  volume  = {20},
  number  = {2},
  pages   = {130--141},
  year    = {1963}
}

@inproceedings{montufar2014number,
  author = {Mont\'{u}far, Guido and Pascanu, Razvan and Cho, Kyunghyun and Bengio, Yoshua},
 booktitle = {Advances in Neural Information Processing Systems},
 editor = {Z. Ghahramani and M. Welling and C. Cortes and N. Lawrence and K. Weinberger},
 pages = {},
 publisher = {Curran Associates, Inc.},
 title = {On the Number of Linear Regions of Deep Neural Networks},
 volume = {27},
 year = {2014}
}

@inproceedings{raghu2017expressive,
  author    = {Raghu, Maithra and Poole, Ben and Kleinberg, Jon and Ganguli, Surya and Sohl-Dickstein, Jascha},
  title     = {On the expressive power of deep neural networks},
  booktitle = {Proc. 34th Int. Conf. Mach. Learn. (ICML)},
  pages     = {2847--2854},
  year      = {2017}
}

@article{roudneff1991cells,
  author  = {Roudneff, Jean-Pierre},
  title   = {Cells with many facets in arrangements of hyperplanes},
  journal = {Discrete Math.},
  volume  = {98},
  number  = {3},
  pages   = {185--191},
  year    = {1991}
}

@inproceedings{serra2018bounding,
  author    = {Serra, Thiago and Tjandraatmadja, Christian and Ramalingam, Srikumar},
  title     = {Bounding and counting linear regions of deep neural networks},
  booktitle = {Proc. Int. Conf. Mach. Learn. (ICML)},
  pages     = {4558--4566},
  year      = {2018}
}

@article{li2025achieving,
  author={Li, Zhiwei and Wang, Cheng},
  journal={IEEE Transactions on Neural Networks and Learning Systems}, 
  title={Achieving Sharp Upper Bounds on the Expressive Power of Neural Networks via Tropical Polynomials}, 
  year={2025},
  volume={36},
  number={2},
  pages={2931-2945},
  doi={10.1109/TNNLS.2024.3350786}
}

@inproceedings{suard2004efficient,
  author    = {Suard, Rapha{\"e}l and L{\"o}fberg, J. and Grieder, Pascal and Kvasnica, Michal and Morari, Manfred},
  title     = {Efficient computation of controller partitions in multi-parametric programming},
  booktitle = {Proc. 43rd IEEE Conference on Decision and Control (CDC)},
  volume    = {4},
  pages     = {3643--3648},
  year      = {2004}
}

@inproceedings{serra2020empirical,
  author    = {Serra, Thiago and Ramalingam, Srikumar},
  title     = {Empirical bounds on linear regions of deep rectifier networks},
  booktitle = {Proc. AAAI Conf. Artif. Intell.},
  volume    = {34},
  number    = {4},
  pages     = {5628--5635},
  year      = {2020}
}

@article{katz2022verification,
  author  = {Katz, Sydney M. and Corso, Anthony L. and Strong, Christopher A. and Kochenderfer, Mykel J.},
  title   = {Verification of image-based neural network controllers using generative models},
  journal = {J. Aerosp. Inf. Syst.},
  volume  = {19},
  number  = {9},
  pages   = {574--584},
  year    = {2022}
}

@inproceedings{ravaioli2022safe,
  author    = {Ravaioli, Umberto J. and Cunningham, James and McCarroll, John and Gangal, Vardaan and Dunlap, Kyle and Hobbs, Kerianne L.},
  title     = {Safe reinforcement learning benchmark environments for aerospace control systems},
  booktitle = {Proc. IEEE Aerosp. Conf. (AERO)},
  pages     = {1--20},
  year      = {2022}
}

@article{brockman2016openai,
  author  = {Brockman, Greg and Cheung, Vicki and Pettersson, Ludwig and Schneider, Jonas and Schulman, John and Tang, Jie and Zaremba, Wojciech},
  title   = {{OpenAI} gym},
  journal = {arXiv preprint arXiv:1606.01540},
  year    = {2016}
}

@inproceedings{he2022characterizing,
  author    = {He, Haoyu and Wei, Tianhao and Zhang, Huan and Liu, Changliu and Tan, Cheng},
  title     = {Characterizing neural network verification for systems with {nn4sysbench}},
  booktitle = {Proc. Workshop Formal Verification Mach. Learn.},
  year      = {2022}
}

@article{brix2023vnncomp_overview,
  author  = {Brix, Christopher and M{\"u}ller, Mark Niklas and Bak, Stanley and Johnson, Taylor T. and Liu, Changliu},
  title   = {First three years of the international verification of neural networks competition ({VNN-COMP})},
  journal = {Int. J. Softw. Tools Technol. Transf.},
  volume  = {25},
  number  = {3},
  pages   = {329--339},
  year    = {2023},
  doi     = {10.1007/s10009-023-00703-4}
}

@inproceedings{zhou2025clip,
  author = {Zhou, Duo and Chavez, Jorge and Chen, Hesun and Hanasusanto, Grani A. and Zhang, Huan},
 booktitle = {Advances in Neural Information Processing Systems},
 doi = {10.52202/085713-5815},
 editor = {D. Belgrave and C. Zhang and H. Lin and R. Pascanu and P. Koniusz and M. Ghassemi and N. Chen},
 pages = {174849--174895},
 publisher = {Curran Associates, Inc.},
 title = {Clip-and-Verify: Linear Constraint-Driven Domain Clipping for Accelerating Neural Network Verification},
 volume = {38, Main Conference},
 year = {2025}
}

@inproceedings{zhang2018efficient,
  author = {Zhang, Huan and Weng, Tsui-Wei and Chen, Pin-Yu and Hsieh, Cho-Jui and Daniel, Luca},
 booktitle = {Advances in Neural Information Processing Systems},
 editor = {S. Bengio and H. Wallach and H. Larochelle and K. Grauman and N. Cesa-Bianchi and R. Garnett},
 pages = {},
 publisher = {Curran Associates, Inc.},
 title = {Efficient Neural Network Robustness Certification with General Activation Functions},
 volume = {31},
 year = {2018}
}

@inproceedings{bak2021nfm,
  author    = {Bak, Stanley},
  title     = {{nnenum}: Verification of {ReLU} neural networks with optimized abstraction refinement},
  booktitle = {Proc. NASA Formal Methods Symp. (NFM)},
  pages     = {19--36},
  year      = {2021}
}

@inproceedings{sun2021learning,
  author    = {Sun, Dawei and Jha, Susmit and Fan, Chuchu},
  title     = {Learning certified control using contraction metric},
  booktitle = {Proc. Conf. Robot Learn. (CoRL)},
  pages     = {1519--1539},
  year      = {2021}
}

@article{owen1998scrambling,
  title={Scrambling {S}obol'and {N}iederreiter--{X}ing Points},
    journal = {Journal of Complexity},
    volume = {14},
    number = {4},
    pages = {466-489},
    year = {1998},
    issn = {0885-064X},
    doi = {https://doi.org/10.1006/jcom.1998.0487},
    url = {https://www.sciencedirect.com/science/article/pii/S0885064X98904873},
    author = {Art B. Owen}
}

@software{bak2026vibecheck,
  author     = {Stanley Bak},
  title      = {{VibeCheck}: Neural Network Verification Tool},
  year       = {2026},
  url        = {https://github.com/stanleybak/vibecheck-nn},
  repository = {https://github.com/stanleybak/vibecheck-nn}
}
